\documentclass[letterpaper, 10 pt, conference]{ieeeconf}
\IEEEoverridecommandlockouts
\usepackage{amsmath,latexsym,amssymb,multicol}
\usepackage[pdftex]{graphicx}
\newcommand{\diag}{\ensuremath{\mathrm{diag}}}

\newtheorem{theorem}{Theorem}
\newtheorem{example}{Example}
\newtheorem{lemma}{Lemma}
\title{\Large \bf On Transfer Function Realizations for Linear Quantum Stochastic Systems}

\author{Symeon Grivopoulos$^1$ and Hendra I. Nurdin$^2$ and Ian R. Petersen$^1$%
\thanks{$^1$ Symeon Grivopoulos and Ian R. Petersen are with the School of Engineering and Information Technology, UNSW Canberra, Canberra BC 2610, Australia {\tt\small symeon.grivopoulos@gmail.com, i.r.petersen@gmail.com}}
\thanks{$^2$ Hendra I. Nurdin is with the School of Electrical Engineering and Telecommunications, UNSW Australia, Sydney NSW 2052, Australia {\tt\small h.nurdin@unsw.edu.au}}
\thanks{This work was supported by the Australian Research Council under grant FL110100020 (Grivopoulos and Petersen) and DP130104191 (Nurdin)}}

\begin{document}

\maketitle

\begin{abstract}
The realization of transfer functions of Linear Quantum Stochastic Systems (LQSSs) is an issue of fundamental importance for the practical applications of such systems, especially as coherent controllers for other quantum systems. In this paper, we review two realization methods proposed by the authors in \cite{pet11,nur10b,nurgripet16, gripet15}. The first one uses a cascade of a static linear quantum-optical network and single-mode optical cavities, while the second uses a feedback network of such cavities, along with static linear quantum-optical networks that pre- and post-process the cavity network inputs and outputs.
\end{abstract}

\section{Introduction}
\label{Introduction}

Linear Quantum Stochastic Systems (LQSSs) are a class of models widely used in linear quantum optics and elsewhere \cite{garzol00,walmil08,wismil10}. In quantum optics, they describe a variety of devices, such as optical cavities, parametric amplifiers, etc., as well as networks of such devices. The mathematical framework for these models is provided by the theory of quantum Wiener processes and the associated Quantum Stochastic Differential Equations \cite{par99,mey95,hudpar84}. Potential applications of linear quantum optics include quantum information and photonic signal processing, see e.g. \cite{niechu00,knilafmil01,ral06,zhajam13,zha14}. Another particularly important application of LQSSs is as coherent quantum feedback controllers for other quantum systems, i.e. controllers that do not perform any measurement on the controlled quantum system, and thus have the potential for increased performance compared to classical controllers, see e.g. \cite{yankim03a,yankim03b,jamnurpet08,nurjampet09,maapet11b,mab08,hammab12,critezsoh13}.

A problem of fundamental importance for applications of LQSSs, is the problem of realization/synthesis: Given a LQSS with specified parameters, how does one engineer that system using basic quantum optical devices, such as optical cavities, parametric amplifiers, phase shifters, beam splitters, squeezers etc.? The synthesis problem comes in two varieties. First, there is the \emph{strict realization} problem which we just described. This type of realization is necessary in the case where the states of the quantum system are meaningful to the application at hand. Examples include quantum information processing algorithms \cite{niechu00,knilafmil01,ral06} and state generation \cite{kog12,mawoopetyam14}. In the case that only the input-output relation of the LQSS is important, we have the problem of \emph{transfer function realization}. This is the case, for example, in controller synthesis \cite{mab08,hammab12,critezsoh13}.

In recent years, solutions have been proposed to both the strict and the transfer function realization problems. For the strict problem, \cite{nurjamdoh09,nur10a} propose a cascade of single-mode cavities realization. This allows for arbitrary couplings of the LQSS to its environment. However, not all possible interactions between cavity modes are possible, because the mode of a cavity can influence only modes of subsequent cavities. For this reason, direct Hamiltonian interactions \cite{nurjamdoh09} and feedback \cite{nur10a} between cavities have been used to ``correct'' the dynamics of the cascade to the desired form. In this article, we review two methods for the transfer function realization of LQSSs, proposed in \cite{pet11,nur10b,nurgripet16, gripet15}. The first method uses a cascade of single-mode cavities. For the case of passive LQSSs, \cite{nur10b} has shown that such a realization is possible for any passive system, in which case all cavities needed to realize it are also passive. The result for the general case is established in \cite{nurgripet16}, where it is shown that a cascade of cavities realization is possible for generic LQSSs. The second method \cite{gripet15} utilizes static linear quantum-optical networks that pre- and post-process the system inputs and outputs, thus leaving a simple ``reduced'' transfer function to be realized. This ``reduced'' transfer function can be realized, in turn, by a concatenation of single-mode cavities in a feedback interconnection through a static linear quantum-optical network. In the case of passive LQSSs, this realization is always possible, and all necessary devices needed for it are also passive.

In the case of passive LQSSs, the realization methods make crucial use of two classic theorems from Linear Algebra, namely Schur's Unitary Triangularization theorem, and the Singular Value Decomposition \cite{horjoh85}, respectively. To extend them from passive LQSSs to general LQSSs that may contain active (quanta producing) quantum optical devices, we prove two analogous matrix factorizations for a class of even-dimensional structured matrices, the so-called doubled-up matrices \cite{goujamnur10,pet10}, in a class of complex spaces with indefinite scalar products, the so-called Krein spaces \cite{gohlanrod83}. Contrary to their classic counterparts, these factorizations do not hold for every doubled-up matrix.

\section{Background Material}
\label{Background Material}

\subsection{Notation and terminology}
\label{Notation and terminology}
\begin{enumerate}
  \item $x^{*}$ denotes the complex conjugate of a complex number $x$ or the adjoint of an operator $x$, respectively. As usual, $\Re x$ and $\Im x$ denote the real and imaginary part of a complex number. The commutator of two operators $X$ and $Y$ is defined as $[X,Y]=XY-YX$.
  \item For a matrix $X=[x_{ij}]$ with number or operator entries, $X^{\#}=[x_{ij}^*]$, $X^{\top}=[x_{ji}]$ is the usual transpose, and $X^{\dag}=(X^{\#})^{\top}$. Also, for a vector $x=[x_i]$ with number or operator entries, we shall use the notation $\check{x}=\bigl(\begin{smallmatrix} x \\ x^{\#} \end{smallmatrix}\bigr)$.
  \item The identity matrix in $n$ dimensions will be denoted by $I_n$, and a $r \times s$ matrix of zeros will be denoted by $0_{r \times s}$. $\delta_{ij}$ denotes the Kronecker delta symbol in $n$ dimensions, i.e. $I_n=[\delta_{ij}]$. $\diag(X_1,X_2,\ldots,X_k)$ is the block-diagonal matrix formed by the square matrices $X_1,X_2,\ldots,X_k$. $[Y_1 Y_2 \ldots Y_k]$ is the horizontal concatenation of the matrices $Y_1, Y_2, \ldots, Y_k$ of equal row dimension.
  \item We define $J_{2k}=\diag(I_k,-I_k)$, and $\Sigma_{2k} = \bigl(\begin{smallmatrix} 0_{k \times k} & I_k \\ I_k & 0_{k \times k} \end{smallmatrix} \bigr)$. We have that $J_{2k}^2=\Sigma_{2k}^2=I_{2k}$ and, $\Sigma_{2k} J_{2k} \Sigma_{2k}=-J_{2k}$. When the dimensions of $I_n$, $0_{r \times s}$, $J_{2k}$ or $\Sigma_{2k}$ can be inferred from context, they will be denoted simply by $I$, $\mathbf{0}$, $J$ and $\Sigma$. Also, $\sigma_2 = \bigl(\begin{smallmatrix} 0 & -\imath \\ \imath & 0 \end{smallmatrix} \bigr)$ is the second Pauli matrix.
  \item We define the \emph{Krein space} ($\mathbb{C}^{2k}$, $J_{2k}$) as the vector space $\mathbb{C}^{2k}$ equipped with the \emph{indefinite inner product} defined by $\langle v,w\rangle_J=v^{\dag}J_{2k}w$, for any $v,w \in \mathbb{C}^{2k}$. The $J$-norm of a vector $v \in \mathbb{C}^{2k}$ is defined by $|v|_J = \sqrt{|\langle v,v\rangle_J|}$, and if it is nonzero, a normalized multiple of $v$ is $v/|v|_J$. For a $2r \times 2s$ matrix $X$ considered as a map from ($\mathbb{C}^{2s}$, $J_{2s}$) to ($\mathbb{C}^{2r}$, $J_{2r}$), its adjoint operator will be called $\flat$-\emph{adjoint} and denoted by $X^{\flat}$, to distinguish it from its usual adjoint $X^{\dag}$. One can show that $X^{\flat}=J_{2s}X^{\dag}J_{2r}$. The $\flat$-adjoint satisfies properties similar to the usual adjoint, namely $(x_1 A + x_2 B)^{\flat}=x_1^* A^{\flat} + x_2^* B^{\flat}$, and $(AB)^{\flat}=B^{\flat}  A^{\flat}$.
  \item Given two $r \times s$ matrices $X_1$, and $X_2$, respectively, we can form the $2r \times 2s$ matrix $X=\bigl(\begin{smallmatrix} X_1 & X_2 \\ X_2^{\#} & X_1^{\#} \end{smallmatrix}\bigr)$. Such a matrix is said to be \emph{doubled-up} \cite{goujamnur10}. It is immediate to see that the set of doubled-up matrices is closed under addition, multiplication and taking ($\flat$-) adjoints. Also, $\Sigma_{2r} X \Sigma_{2s}=X^{\#}$, if and only if $X^{2r \times 2s}$ is doubled-up When referring to a doubled-up matrix $X^{2r \times 2s}$, $X_1^{r \times s}$ and $X_2^{r \times s}$, will denote its upper-left and upper-right blocks.
  \item A $2k \times 2k$ complex matrix R is called \emph{Bogoliubov} if it is doubled-up and $\flat$-unitary, i.e $RR^{\flat}=R^{\flat}R=I_{2m}$. The set of these matrices forms a non-compact Lie group known as the Bogoliubov group. Bogoliubov matrices are isometries of Krein spaces.
\end{enumerate}

\subsection{Linear Quantum Stochastic Systems}
\label{Linear Quantum Stochastic Systems}

The material in this subsection is fairly standard, and our presentation aims mostly at establishing notation and terminology. To this end, we follow the review paper \cite{pet10}. For the mathematical background necessary for a precise discussion of LQSSs, some standard references are \cite{par99,mey95,hudpar84}, while for a Physics perspective, see \cite{garzol00,garcol85}. The references \cite{nurjamdoh09,edwbel05,goujam09,gougohyan08,goujamnur10} contain a lot of relevant material, as well.

The systems we consider in this work are collections of quantum harmonic oscillators interacting among themselves, as well as with their environment. The $i$-th harmonic oscillator ($i=1,\ldots,n$) is described by its position and momentum variables, $x_i$ and $p_i$, respectively. These are self-adjoint operators satisfying the \emph{Canonical Commutation Relations} (CCRs) $[x_i,x_j]=0$, $[p_i,p_j]=0$, and $[x_i,p_j]=\imath\delta_{ij}$, for $i,j=1,\ldots,n$. We find it more convenient to work with the so-called annihilation and creation operators $a_i=\frac{1}{\sqrt{2}}(x_i + \imath p_i)$, and $a_i^*=\frac{1}{\sqrt{2}}(x_i - \imath p_i)$. They satisfy the CCRs $[a_i,a_j]=0$, $[a_i^*,a_j^*]=0$, and $[a_i,a_j^*]=\delta_{ij}$, $i,j=1,\ldots,n$. In the following, $a=(a_1,a_2,\ldots,a_n)^{\top}$.

The environment is modelled as a collection of bosonic heat reservoirs. The $i$-th heat reservoir ($i=1,\ldots,m$) is described by the bosonic field annihilation and creation operators $\mathcal{A}_i(t)$ and $\mathcal{A}_i^*(t)$, respectively. The field operators are \emph{adapted quantum stochastic processes} with forward differentials $d\mathcal{A}_i(t)= \mathcal{A}_i(t+dt)-\mathcal{A}_i(t)$, and $d\mathcal{A}_i^*(t)= \mathcal{A}_i^*(t+dt)-\mathcal{A}_i^*(t)$. They satisfy the quantum It\^{o} products $d\mathcal{A}_i(t) d\mathcal{A}_j(t)=0$, $d\mathcal{A}_i^*(t) d\mathcal{A}_j^*(t)=0$, $d\mathcal{A}_i^*(t) d\mathcal{A}_j(t)=0$, and $d\mathcal{A}_i(t) d\mathcal{A}_j^*(t)=\delta_{ij} dt$. In the following, $\mathcal{A}=(\mathcal{A}_1,\mathcal{A}_2,\ldots,\mathcal{A}_m)^{\top}$.

To describe the dynamics of the harmonic oscillators and the quantum fields (noises), we need to introduce certain operators. We begin with the class of \emph{annihilator only} LQSSs. We also refer to such systems as \emph{passive} LQSSs, because systems in this class describe optical devices such as damped optical cavities, that do not require an external source of quanta for their operation. First, we have the Hamiltonian operator $H=a^{\dag}Ma$, which specifies the dynamics of the harmonic oscillators in the absence of any environmental influence. $M$ is a $n \times n$ Hermitian matrix referred to as the Hamiltonian matrix. Next, we have the coupling operator $L$ (vector of operators) that specifies the interaction of the harmonic oscillators with the quantum fields. $L$ depends linearly on the annihilation operators, and can be expressed  as $L=Na$. $N$ is called the coupling matrix. Finally, we have the unitary scattering matrix $S^{m \times m}$, that describes the interactions between the quantum fields themselves. In practice, it represents the unitary transformation effected on the heat reservoir modes by a static passive linear optical network that precedes the LQSS, see Subsection \ref{Static Linear Optical Devices and Networks}.

In the \emph{Heisenberg picture} of Quantum Mechanics, the joint evolution of the harmonic oscillators and the quantum fields is described by the following system of \emph{Quantum Stochastic Differential Equations} (QSDEs):
\begin{eqnarray}
da &=& \Big(-\imath M -\frac{1}{2}N^{\dag}N \Big) \,a\, dt -N^{\dag}S\, d\mathcal{A}, \nonumber \\
d\mathcal{A}_{out} &=& N a\, dt + S\, d\mathcal{A}. \label{Passive LQSS 1}
\end{eqnarray}
The field operators $\mathcal{A}_{i \, out}(t), i=1,\ldots,m$, describe the outputs of the system. We can generalize (\ref{Passive LQSS 1}) by allowing the system inputs to be not just quantum noises, but to contain a ``signal part'', as well. Such is the case when the output of a passive LQSS is fed into another passive LQSS. So we substitute the more general input and output notations $\mathcal{U}$ and $\mathcal{Y}$, for $\mathcal{A}$ and $\mathcal{A}_{out}$, respectively. The forward differentials $d\mathcal{U}$ and $d\mathcal{Y}$ of $m$-dimensional inputs and outputs, respectively, contain quantum noises, as well as linear combinations of variables of other systems. The resulting QSDEs are the following:
\begin{eqnarray}
da &=& \Big(-\imath M -\frac{1}{2}N^{\dag}N \Big) \,a\, dt -N^{\dag}S\, d\mathcal{U}, \nonumber \\
d\mathcal{Y} &=& N a\, dt + S\, d\mathcal{U}. \label{Passive LQSS 2}
\end{eqnarray}
One can show that the structure of (\ref{Passive LQSS 2}) is preserved under linear transformations of the state $\hat{a}= V a$, if and only if $V$ is unitary. Under such a state transformation, the system parameters $(S,N,M)$ transform according to $(\hat{S},\hat{N},\hat{M})=(S,NV^{-1},VMV^{\dag})$. From the point of view of Quantum Mechanics, $V$ must be unitary so that the new annihilation and creation operators satisfy the correct CCRs.

General LQSSs may contain \emph{active} devices that require an external source of quanta for their operation, such as degenerate parametric amplifiers. In this case, system and field creation operators appear in the QSDEs for system and field annihilation operators, and vice versa. Since these are adjoint operators which have to be treated as separate variables, this leads to the appearance of doubled-up matrices in the corresponding QSDEs. To describe the most general linear dynamics of harmonic oscillators and quantum noises, we introduce generalized versions of the Hamiltonian operator, the coupling operator, and the scattering matrix defined above. We begin with the Hamiltonian operator
\begin{eqnarray*}
H=\frac{1}{2} \left(\begin{array}{l} a \\ a^{\#} \\ \end{array}\right)^{\dag}
\left(\begin{array}{cc} M_1 & M_2 \\ M_2^{\#} & M_1^{\#} \\ \end{array} \right) \left(\begin{array}{l} a \\ a^{\#} \\ \end{array}\right) =\frac{1}{2}\check{a}^{\dag}M\check{a},
\end{eqnarray*}
which specifies the dynamics of the harmonic oscillators in the absence of any environmental influence. The $2n \times 2n$ Hamiltonian matrix $M$ is Hermitian and doubled-up. Next, we have the coupling operator $L$ (vector of operators) that specifies the interaction of the harmonic oscillators with the quantum fields. $L$ depends linearly on the creation and annihilation operators, $L=N_1 a+N_2 a^{\#}$. We construct the doubled-up coupling matrix $N^{2m \times 2n}$ from $N_1^{m \times n}$ and $N_2^{m \times n}$. Finally, we have the Bogoliubov generalized scattering matrix $S^{2m \times 2m}$, that describes the interactions between the quantum fields themselves. In practice, it represents the Bogoliubov transformation effected on the heat reservoir modes by a general static linear quantum optical network that precedes the LQSS, see Subsection \ref{Static Linear Optical Devices and Networks}, and \cite{goujamnur10}.

In the \emph{Heisenberg picture} of Quantum Mechanics, the joint evolution of the harmonic oscillators and the quantum fields is described by the following system of \emph{Quantum Stochastic Differential Equations} (QSDEs):
\begin{eqnarray}
d\check{a} &=& \big(-\imath JM -\frac{1}{2}N^{\flat}N\big)\, \check{a} dt - N^{\flat} S d\check{\mathcal{U}}, \nonumber \\
d\check{\mathcal{Y}} &=& N \check{a} dt + S d\check{\mathcal{U}}. \label{General LQSS2}
\end{eqnarray}
The forward differentials $d\mathcal{U}$ and $d\mathcal{Y}$ of $m$-dimensional inputs and outputs, respectively, contain quantum noises, as well as a signal part (linear combinations of variables of other systems). One can show that the structure of (\ref{General LQSS2}) is preserved under linear transformations of the state $\check{\tilde{a}} = V \check{a}$, if and only if $V$ is Bogoliubov. In that case the system parameters $(S,N,M)$ transform according to $(\tilde{S},\tilde{N},\tilde{M})= (S,NV^{-1},(V^{-1})^{\dag}MV^{-1})$. From the point of view of Quantum Mechanics, $V$ must be Bogoliubov so that the new annihilation and creation operators satisfy the correct CCRs.

\begin{figure}[!h]
\begin{center}
\scalebox{.3}{\includegraphics{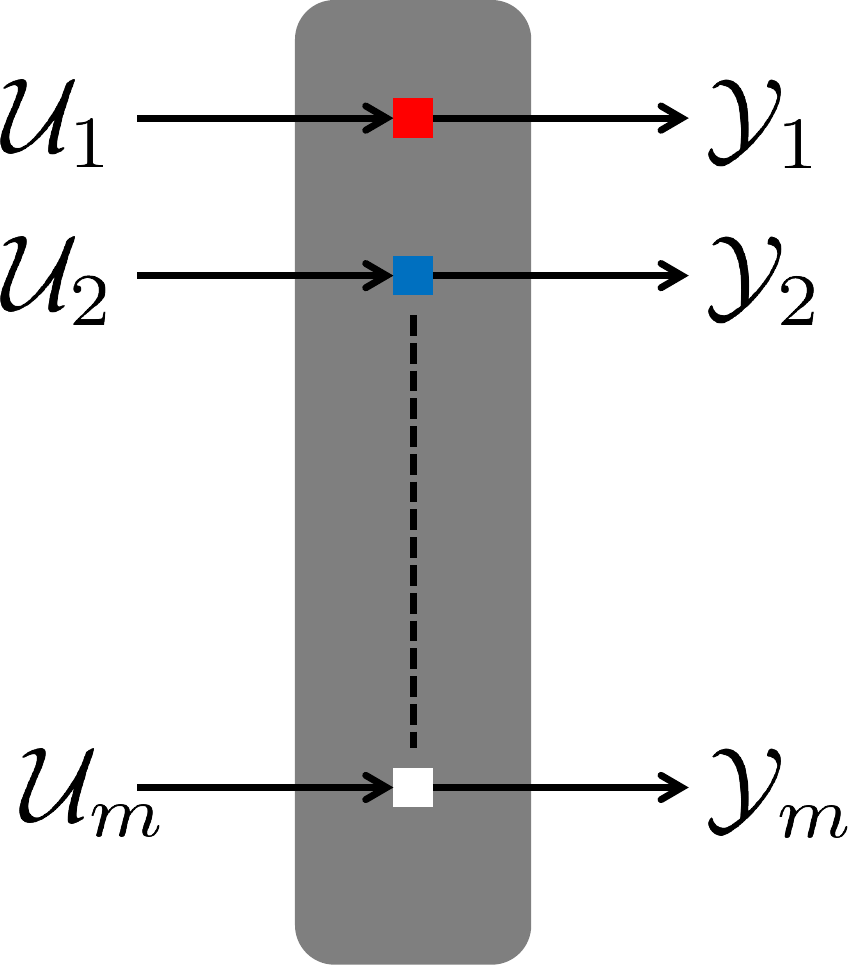}} \caption{Graphical representation of a multi-port cavity. The gray block represents the cavity, and the small squares represent ports. Red is used for passive ports, blue for active ports, and white for all other cases.} \label{General_multiport_cavity}
\end{center}
\end{figure}

We end this subsection with the model of the  single-mode optical cavity, which is the basic device for the proposed realization methods in this paper. It is described by its optical mode $a$, with Hamiltonian matrix $M=\diag(\Delta,\Delta)$, where $\Delta \in \mathbb{R}$ is the so-called cavity detuning. For a cavity with $m$ inputs/outputs, we let $N_1=(e^{\imath\phi_1}\, \sqrt{\kappa_1},\ldots,e^{\imath\phi_m}\, \sqrt{\kappa_m})^{\top}$, and $N_2=(e^{\imath\theta_1}\, \sqrt{g_1},\ldots,e^{\imath\theta_m}\, \sqrt{g_m})^{\top}$. $\kappa_i$ and $g_i$ will be called the \emph{passive} and the \emph{active coupling coefficient} of the $i$-th quantum noise to the cavity, respectively. When $g_i=0$, the interaction of the cavity mode with the $i$-th quantum noise will be referred to as \emph{(purely) passive}, and when $\kappa_i=0$, it will be referred to as \emph{(purely) active}. The model of a cavity with $m$ inputs/outputs, is the following:
\begin{eqnarray}
da &=& \Big[-\imath\Delta -\frac{1}{2}\big(N_1^{\dag}N_1 -N_2^{\top}N_2^{\#}\big)\Big] a\, dt \nonumber \\
&-& N_1^{\dag}d\mathcal{U} + N_2^{\top} d\mathcal{U}^{\#} \nonumber \\
&=&\Big( -\imath\Delta -\frac{\gamma}{2} \Big) a\, dt \nonumber \\
&+&\sum_{i=1}^{m} \Big[-e^{-\imath\phi_i}\,\sqrt{\kappa_i}\, d\mathcal{U}_i + e^{\imath\theta_i}\, \sqrt{g_i}\, d\mathcal{U}_i^{\#} \Big], \nonumber \\
d\mathcal{Y}_i &=& e^{\imath\phi_i}\, \sqrt{\kappa_i}\, a \,dt + e^{\imath\theta_i}\, \sqrt{g_i}\, a^{\#} \,dt + d\mathcal{U}_i,  \label{General cavity model}
\end{eqnarray}
$i=1,\ldots,m$, where $\gamma=\sum_{i=1}^{m}(\kappa_i - g_i)$. If a quantum noise couples passively to the cavity, the corresponding interaction may be realized with a partially transmitting  mirror. For an interaction that has an active component, a more complicated implementation is needed, which makes use of an auxiliary cavity, see e.g. \cite{nurjamdoh09} for the details. From now on, we shall use the system-theoretic term \emph{port} for any part of the experimental set-up that realizes an interaction of the cavity mode with a quantum noise (where an input enters and an output exits the cavity). Figure \ref{General_multiport_cavity} is a graphical representation of a  multi-port cavity modelled by equations (\ref{General cavity model}).

\subsection{Static Linear Optical Devices and Networks}
\label{Static Linear Optical Devices and Networks}

\begin{figure}[!h]
\begin{center}
\scalebox{.3}{\includegraphics{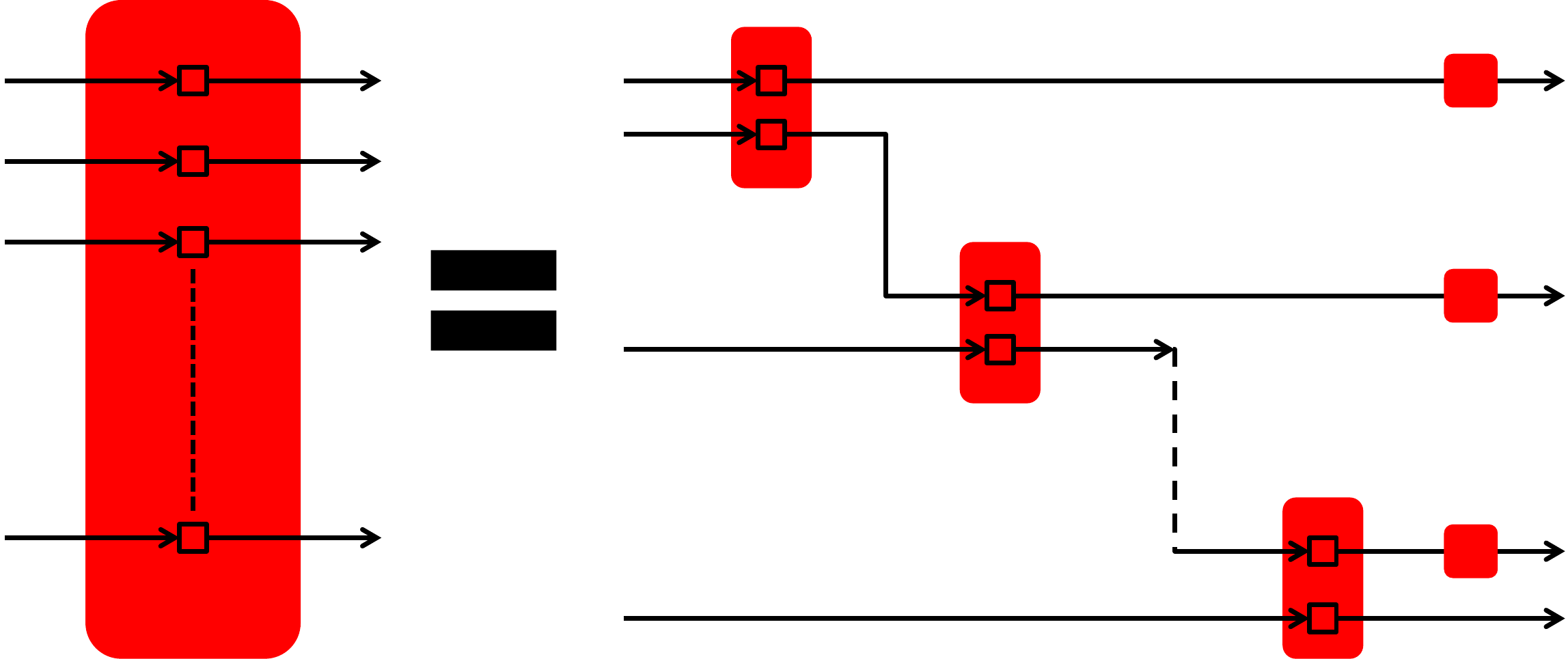}} \caption{Graphical representation of a passive network and its decomposition in terms of beam splitters and phase shifters. In subsequent figures, red blocks will always represent passive static devices and networks.} \label{Passive_network_decomposition}
\end{center}
\end{figure}

\begin{figure}[!h]
\begin{center}
\scalebox{.3}{\includegraphics{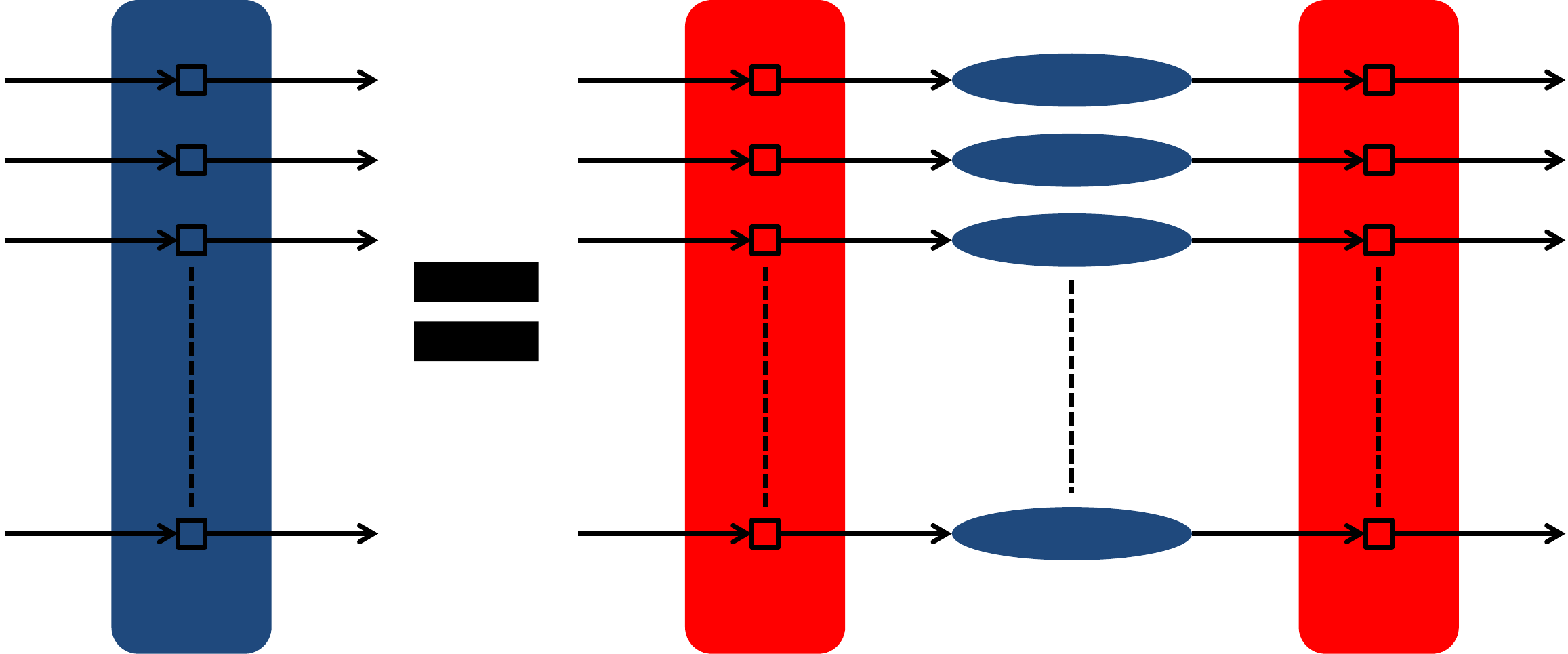}} \caption{Graphical representation of an active network and its decomposition in terms of passive networks and squeezers. In subsequent figures, blue blocks will always represent active static devices and networks.} \label{General_network_decomposition}
\end{center}
\end{figure}

Besides the single-mode cavities discussed above, the proposed realization methods make use of static linear quantum optical devices and networks, as well. Useful references for this material are  \cite{leo03,nurjamdoh09, leoneu04, bra05}. The most basic such devices are the following:
\begin{enumerate}
  \item \textbf{The phase shifter:} This device produces a phase shift in its input optical field. That is, if $\mathcal{U}$ and $\mathcal{Y}$ are its input and output fields, respectively, then $\mathcal{Y}=e^{\imath\theta}\,\mathcal{U}$. Notice that $\mathcal{Y}^* \mathcal{Y}=\mathcal{U}^* \mathcal{U}$. This means that the energy of the output field is equal to that of the input field, and hence the device is passive.
  \item \textbf{The beam splitter:} This device produces linear combinations of its two input fields. If we denote its inputs by $\mathcal{U}_1$ and $\mathcal{U}_2$, and its outputs by $\mathcal{Y}_1$ and $\mathcal{Y}_2$, then
\[ \left(\begin{array}{c}
\mathcal{Y}_1 \\
\mathcal{Y}_2 \\
\end{array}\right)= R \left(\begin{array}{c}
\mathcal{U}_1 \\
\mathcal{U}_2 \\
\end{array}\right),  \]
where
\[ R=e^{\imath \zeta} \left(\begin{array}{rr}
e^{\imath \frac{\phi + \psi}{2}}\,\cos\frac{\theta}{2} & e^{\imath \frac{\psi-\phi}{2}}\,\sin\frac{\theta}{2} \\[.25em]
-e^{\imath \frac{\phi - \psi}{2}}\,\sin\frac{\theta}{2} & e^{-\imath \frac{\phi + \psi}{2}}\,\cos\frac{\theta}{2} \\
\end{array}\right). \]
$\theta$ is called the mixing angle of the beam splitter. $\phi$ and $\psi$ are phase differences in the two input and the two output fields, respectively, produced by phase shifters. $\zeta$ is a common phase shift in both output fields. This form of $R$ corresponds to a general $2 \times 2$ unitary matrix. Because $R \in U(2)$, we can see that
\begin{eqnarray*}
&&\left(\begin{array}{cc}
\mathcal{Y}_1^* & \mathcal{Y}_2^* \\
\end{array}\right)\left(\begin{array}{c}
\mathcal{Y}_1 \\
\mathcal{Y}_2 \\
\end{array}\right) \\
&=&\left(\begin{array}{cc}
\mathcal{U}_1^* & \mathcal{U}_2^* \\
\end{array}\right) R^{\dag}R \left(\begin{array}{c}
\mathcal{U}_1 \\
\mathcal{U}_2 \\
\end{array}\right) \\
&=&\left(\begin{array}{cc}
\mathcal{U}_1^* & \mathcal{U}_2^* \\
\end{array}\right)\left(\begin{array}{c}
\mathcal{U}_1 \\
\mathcal{U}_2 \\
\end{array}\right),
\end{eqnarray*}
and hence the total energy of the output fields is equal to that of the input fields.
  \item \textbf{The squeezer:} This device reduces the variance in the real quadrature $\mathcal{(U+U^*)}/2$, or the imaginary quadrature $\mathcal{(U-U^*)}/2\imath$ of an input field $\mathcal{U}$, while increasing the variance in the other. Its operation is described by
\[ \left(\begin{array}{c}
\mathcal{Y} \\
\mathcal{Y}^* \\
\end{array}\right)= R \left(\begin{array}{c}
\mathcal{U} \\
\mathcal{U}^* \\
\end{array}\right),  \]
where
\[ R=\left(\begin{array}{cc}
e^{\imath (\phi + \psi)}\,\cosh x & e^{\imath (\psi - \phi)}\,\sinh x \\[.25em]
e^{\imath (\phi - \psi)}\,\sinh x & e^{-\imath (\phi + \psi)}\,\cosh x \\\end{array}\right). \]
$x \in \mathbb{R}$ is the squeezing parameter, and $\phi,\psi$ are phase shifts in the input and the output field, respectively, produced by phase shifters. This form of $R$ represents a general $2 \times 2$ Bogoliubov matrix. It is easy to show that $\mathcal{Y}^* \mathcal{Y} \neq \mathcal{U}^* \mathcal{U}$, for $x\neq 0$, and hence energy is not conserved. So, the squeezer is an active device.
\end{enumerate}
By connecting various static linear optical devices, we may form static linear optical networks (multi-port devices). When a network is composed solely of passive devices, it is called passive. The input-output relation of a passive static network with $m$ inputs and outputs, $\mathcal{U}= (\mathcal{U}_1,\ldots,\mathcal{U}_m)^{\top}$ and $\mathcal{Y}=(\mathcal{Y}_1,\ldots,\mathcal{Y}_m)^{\top}$, respectively, is $\mathcal{Y}=R\,\mathcal{U}$, with $R\in U(m)$. Such a network is a multi-dimensional generalization of the beam splitter and is sometimes called a multi-beam splitter. It turns out that any passive static network can be constructed exclusively from phase shifters and beam splitters \cite{reczeiber94}. This is due to the fact that an $m \times m$ unitary matrix can be factorized in terms of matrices representing either phase shifting of an optical field in the network or beam splitting between two optical fields in the network, see Figure \ref{Passive_network_decomposition}.
In the case of general static networks that may include active devices, the input-output relation takes the form $\check{\mathcal{Y}}= R \check{\mathcal{U}}$, where $R$ is a $2m \times 2m$ Bogoliubov matrix. For every Bogoliubov matrix, the following factorization holds:
\[ R= \left(\begin{array}{cc}
U_2 & \mathbf{0} \\
\mathbf{0} & U_2^{\#} \\
\end{array}\right) \left(\begin{array}{cc}
\cosh X & \sinh X \\
\sinh X & \cosh X \\
\end{array}\right) \left(\begin{array}{cc}
U_1 & \mathbf{0} \\
\mathbf{0} & U_1^{\#} \\
\end{array}\right),
 \]
where $U_1, U_2 \in U(m)$ and $X=\diag(x_1,x_2,\ldots,x_m)$, with $x_i \in \mathbb{R}, i=1,\ldots,m$. This factorization is known as Bloch-Messiah reduction \cite{nurjamdoh09,leoneu04,bra05}. The physical interpretation of this equation is that a general static network may be implemented as a sequence of three static networks: First comes a passive static network (multi-beam splitter) implementing the unitary transformation $U_1$. Then follows an active static network made of $m$ squeezers, each acting on an output of the first network, and finally, the outputs of the squeezers are fed into a second multi-beam splitter implementing the unitary transformation $U_2$. This is depicted in Figure \ref{General_network_decomposition}. Because of this structure, a general static network is sometimes called a multi-squeezer.

\section{Realizations of Passive Linear Quantum Stochastic Systems}
\label{Realizations of Passive Linear Quantum Stochastic Systems}

In this section, we present two transfer function realization methods for the case of passive LQSSs. Besides the importance of passive LQSSs in applications, they offer the simplest context in which to present the methods.

\subsection{Cascade Realization}
\label{Cascade Realization, Passive Case}

\begin{figure}[!h]
\begin{center}
\scalebox{.4}{\includegraphics{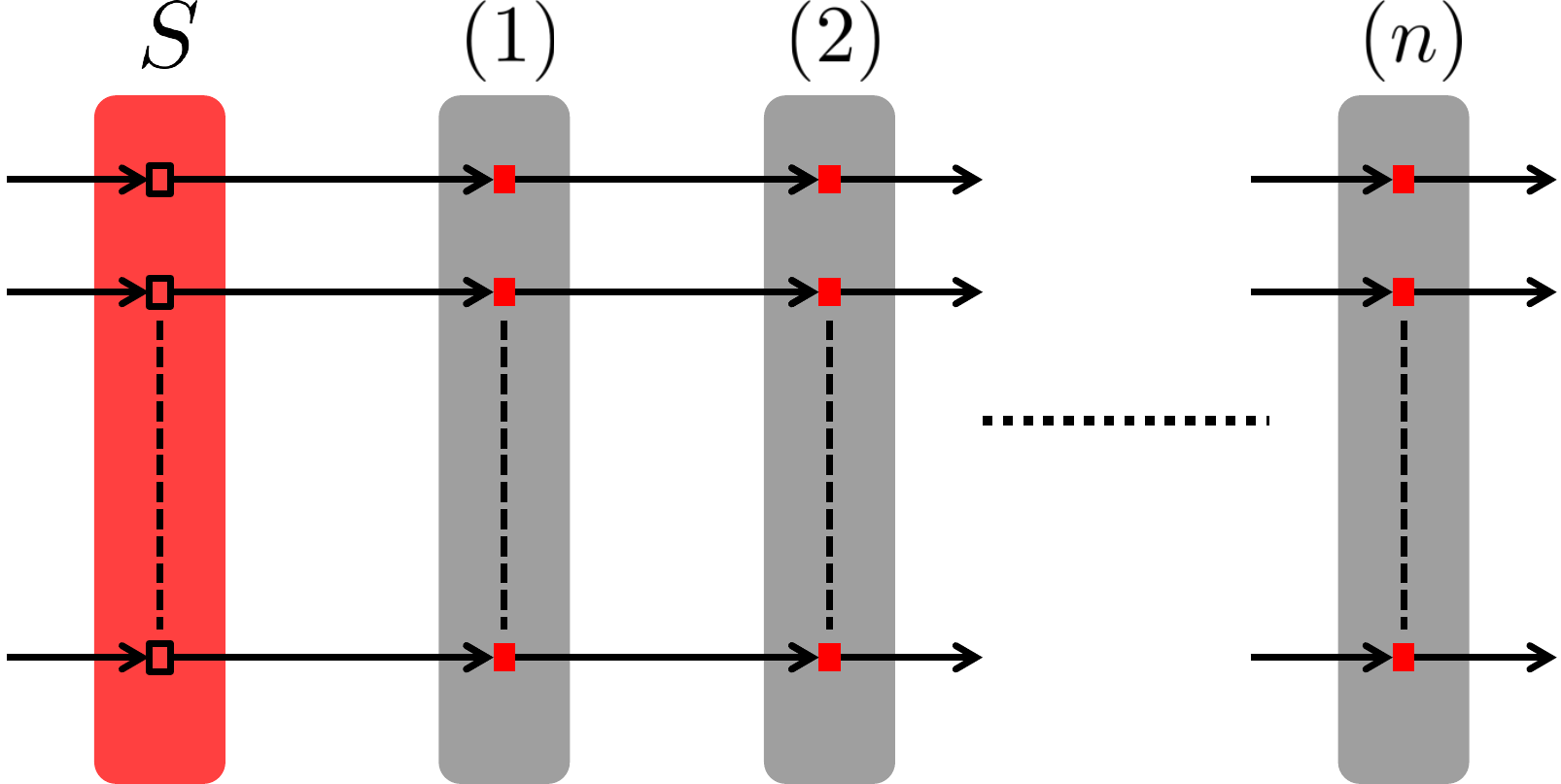}} \caption{Graphical representation of the cascade realization method for passive LQSS transfer functions.} \label{Passive_Cascade}
\end{center}
\end{figure}

We begin with the cavity cascade realization previously obtained in \cite{nur10b} using the real quadrature form of a LQSS (position-momentum operators). Here, we present this cascade realization using a complex formalism (creation-annihilation operators) that simplifies the proof considerably, see also \cite[Subsection 4.1]{gouzha15}. We present this method in the following theorem:
\begin{theorem}\label{Cascade realization, passive case thm}
Given a passive linear quantum stochastic system with Hamiltonian matrix $M^{n \times n}$, coupling operator $N^{m \times n}$, and scattering matrix $S^{m \times m}$, its transfer function can be realized by the following cascade of a multi-beam splitter and $n$ $m$-port passive cavities:
\begin{eqnarray}
\mathcal{Y}_{(0)} &=& S\mathcal{U}, \nonumber \\
\mathcal{U}_{(1)} &=& \mathcal{Y}_{(0)}, \nonumber \\
da_1 &=& \Big(-\imath M_{(1)} -\frac{1}{2}N_{(1)}^{\dag}N_{(1)} \Big) \,a_1 dt -N_{(1)}^{\dag}\, d\mathcal{U}_{(1)}, \nonumber \\
d\mathcal{Y}_{(1)} &=& N_{(1)} a_1 dt + d\mathcal{U}_{(1)}, \nonumber \\
\mathcal{U}_{(2)} &=& \mathcal{Y}_{(1)}, \nonumber \\
da_2 &=& \Big(-\imath M_{(2)} -\frac{1}{2}N_{(2)}^{\dag}N_{(2)} \Big) \,a_2 dt - N_{(2)}^{\dag}\, d\mathcal{U}_{(2)}, \nonumber \\
d\mathcal{Y}_{(2)} &=& N_{(2)} a_2 dt + d\mathcal{U}_{(2)}, \nonumber \\
&\vdots& \nonumber \\
\mathcal{U}_{(n)} &=& \mathcal{Y}_{n-1}, \nonumber \\
da_n &=& \Big(-\imath M_{(n)} -\frac{1}{2}N_{(n)}^{\dag}N_{(n)} \Big) \,a_n dt -N_{(n)}^{\dag}\, d\mathcal{U}_{(n)}, \nonumber \\
d\mathcal{Y}_{(n)} &=& N_{(n)} a_n dt + d\mathcal{U}_{(n)}, \nonumber \\
\mathcal{Y} &=& \mathcal{Y}_{(n)}. \label{Cascade realization, passive case eqn}
\end{eqnarray}
The cavity parameters $M_{(i)} \in \mathbb{R}$, and $N_{(i)} \in \mathbb{C}^m$, $i=1,\ldots,n$, are determined as follows: Define $F=-\imath M -\frac{1}{2}N^{\dag}N$, and let $V$ be a unitary matrix such that $VFV^{\dag}$ is lower-triangular. Then, $M_{(i)}=-\Im(VFV^{\dag})_{ii}$, and $[N_{(1)} N_{(2)} \ldots N_{(n)}]=NV^{\dag}$. $\square$
\end{theorem}
Figure \ref{Passive_Cascade} provides a graphical representation of the cascade realization method of Theorem \ref{Cascade realization, passive case thm}.

\noindent\textbf{Proof:} It is a straightforward calculation to show that the cascade system (\ref{Cascade realization, passive case eqn}) can be put in the following form:
\begin{eqnarray*}
da &=& \hat{F}\,a\, dt -\hat{N}^{\dag}S\, d\mathcal{U}, \\
d\mathcal{Y} &=& \hat{N} a\, dt + S\, d\mathcal{U},
\end{eqnarray*}
where $a=(a_1,\ldots,a_n)^{\top}$, $\hat{N}=[N_{(1)} N_{(2)} \ldots N_{(n)}]$, and
\setlength{\arraycolsep}{-3pt}
\begin{eqnarray*}
\hat{F}= \hspace{22em} \\
\left(\begin{array}{cccc}
-\imath M_{(1)} \! -\frac{1}{2}N_{(1)}^{\dag}N_{(1)}  &  & \text{\Huge0} &  \\
-N_{(2)}^{\dag} N_{(1)} & -\imath M_{(2)} \! -\frac{1}{2}N_{(2)}^{\dag}N_{(2)} &  &  \\
-N_3^{\dag} N_{(1)} & -N_3^{\dag} N_{(2)} & -\imath M_3 \! -\frac{1}{2}N_3^{\dag}N_3 & \\
\vdots & \vdots & \vdots & \ddots \\
\end{array}\right)
\end{eqnarray*}
\setlength{\arraycolsep}{3pt}
is lower-triangular. Now, given a passive LQSS with parameters $(M,N,S)$, let $F=-\imath M -\frac{1}{2}N^{\dag}N$. From Schur's Unitary Triangularization theorem \cite{horjoh85}, there exists a unitary $V$ such that $VFV^{\dag}=VFV^{-1}$ is lower-triangular. Using $V$ as a state transformation, we obtain a realization of the system dynamics in cascade form. The parameters of the cascade realization are given in terms of the original parameters by $[N_{(1)} N_{(2)} \ldots N_{(n)}]=\hat{N}=NV^{-1}=NV^{\dag}$, and $M_{(i)}=-\Im\hat{F}_{ii}, i=1,\ldots,n$, where $\hat{F}=VFV^{\dag}=VFV^{-1}$. Since the transfer function of a linear system is independent of its particular realization, it follows that the transfer function of a passive LQSS can always be realized by the cascade form given in the theorem.$\blacksquare$

We should point out that this realization is not unique, but depends on the order of appearance of the eigenvalues of $F$ on the diagonal of its lower-triangular form $\hat{F}$, which leads to different $\hat{F}$'s and $V$'s. We demonstrate this method with an illustrative example.
\begin{example}\label{example 1}
Consider the 3-mode, 3-input passive linear quantum stochastic system with the following parameters:
\begin{equation*}
M = \left(\begin{array}{rrr}
5&1&-2 \\
1&3&0 \\
-2&0&4 \\
\end{array}\right), \
N = \left(\begin{array}{rrr}
1&2&1 \\
0&-1&3 \\
2&3&5 \\
\end{array}\right),
\end{equation*}
and $S=I_3$. We compute $F=-\imath M -\frac{1}{2}N^{\dag}N$ to be equal to
\begin{eqnarray*}
F=\left(\begin{array}{ccc}
-2.5 - 5\imath & -4 -\imath & -5.5 + 2\imath \\
-4 - \imath & -7 - 3\imath & -7 \\
-5.5 + 2\imath & -7 & -17.5 - 4\imath \\
\end{array}\right).
\end{eqnarray*}
We compute a lower-triangular $\hat{F}$ and a unitary $V$, such that $\hat{F}=VFV^{\dag}$:
\begin{eqnarray*}
V &=&\left(\begin{array}{rrr}
 -0.2960 &  -0.4326 &  -0.8168 \\
 -0.7411 &  -0.2377 &   0.3895 \\
 -0.2394 &  -0.3857 &   0.3169 \\
\end{array}\right) \\
&+& \imath\left(\begin{array}{rrr}
 0.1021 &  0.0955 &  0.1963 \\
 -0.1810 &  0.4365 &  -0.1388 \\
 0.5125 & -0.6387  &  0.1514 \\
\end{array}\right), \\
\hat{F} &=& \left(\begin{array}{rrr}
-23.1603 &    0     &     0 \\
-0.4997  & -1.9103  &     0 \\
0.9734  &  -1.1608  & -1.9294 \\
\end{array}\right) \\
&+& \imath \left(\begin{array}{rrr}
-3.1301 &     0    &    0 \\
0.8117  &  -5.5835 &    0 \\
0.4336  & -3.6141  & -3.2865 \\
\end{array}\right).
\end{eqnarray*}
The parameters $M_{(1)}$,$M_{(2)}$,$M_{(3)} \in \mathbb{R}$, and $N_{(1)}$,$N_{(2)}$,$N_{(3)} \in \mathbb{C}^3$ are given by $(M_{(1)},M_{(2)},M_{(3)})=-\Im(\hat{F}_{11},\hat{F}_{22},\hat{F}_{33})=(3.1301,5.5835,3.2865)$, and
\begin{eqnarray*}
&& [N_{(1)}\ N_{(2)}\ N_{(3)}]=NV^{\dag} \\
&=& \left(\begin{array}{rrr}
 -1.9781 &  -0.8270 &  -0.6940\\
 -2.0177 &  1.4064  &  1.3364\\
 -5.9738 & -0.2476  & -0.0517\\
\end{array}\right) \\
&+& \imath \left(\begin{array}{rrr}
 -0.4894 &  -0.5531  &  0.6135\\
 -0.4935 &   0.8529  & -1.0928\\
 -1.4722 &  -0.2534  &  0.1342\\
\end{array}\right). \square
\end{eqnarray*}
\end{example}

\subsection{Realization Using Static Networks for Input/Output Processing and Feedback}
\label{Realization Using Static Networks for Input/Output Processing and Feedback, Passive Case}

\begin{figure}[!h]
\begin{center}
\scalebox{.25}{\includegraphics{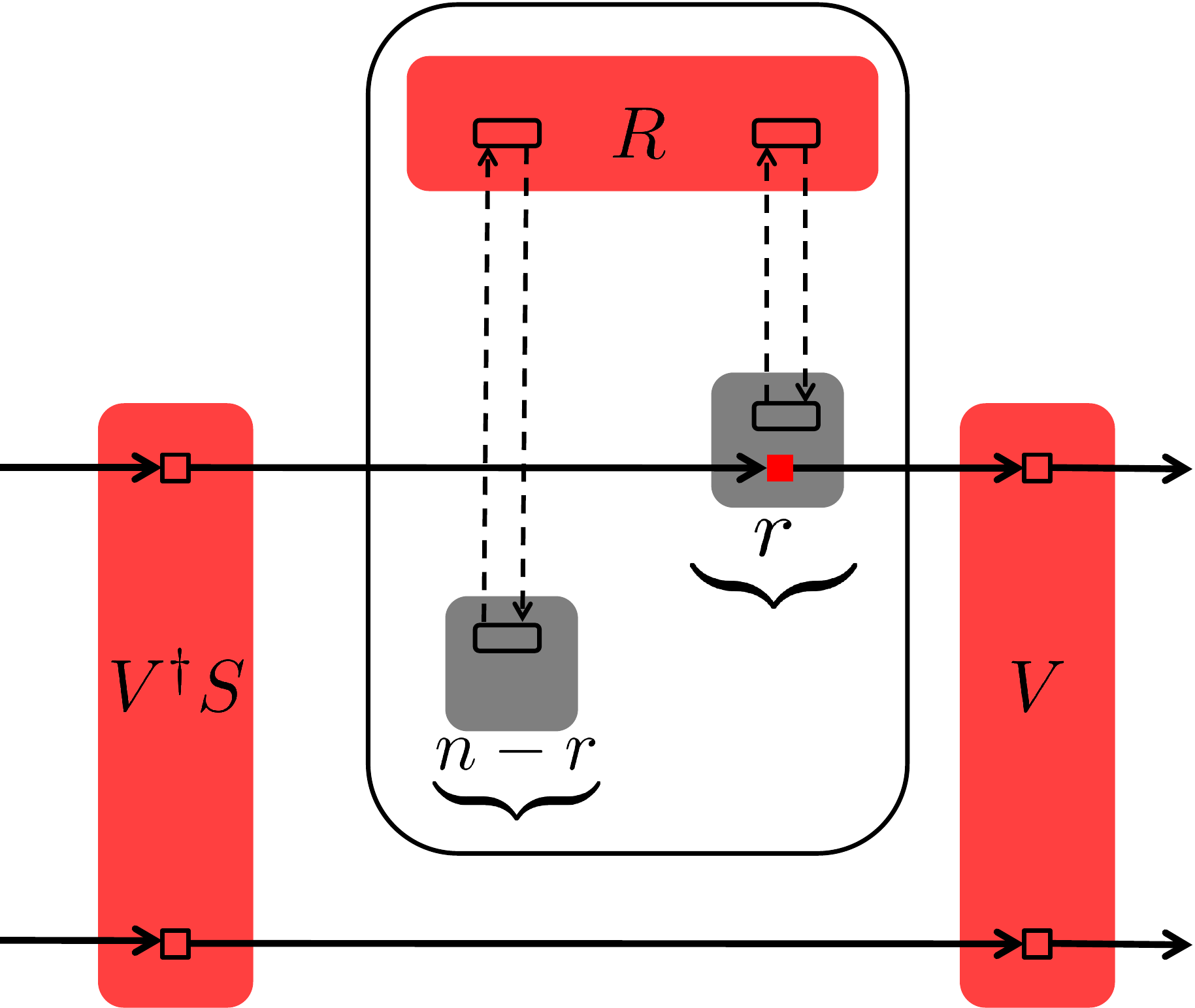}} \caption{Graphical representation of the realization method for passive LQSS transfer functions of Theorem \ref{Feedback network realization, passive case thm}. Each cavity represents all others of its type.} \label{Passive_Network_3}
\end{center}
\end{figure}

Next, we present the realization method of \cite{gripet15} for the case of passive LQSSs, in the following theorem:
\begin{theorem}\label{Feedback network realization, passive case thm}
Given a passive linear quantum stochastic system with Hamiltonian matrix $M^{n \times n}$, coupling operator $N^{m \times n}$, and scattering matrix $S^{m \times m}$, let
\[ G(s)=\Big[ I-N\, \Big(sI+\imath M+\frac{1}{2}\, N^{\dag}N\Big)^{-1}N^{\dag}\Big] S \]
be its transfer function. Let $N=V\hat{N}W^{\dag}$ be the singular value decomposition of the coupling matrix $N$, with
\begin{equation}\label{N_hat, passive case}
\hat{N} = \left(\begin{array}{ccc|c}
\sqrt{\kappa}_1 & & & \\
& \ddots & & \mathbf{0} \\
& & \sqrt{\kappa}_r & \\ \hline
& \mathbf{0} & & \mathbf{0} \\
\end{array} \right).
\end{equation}
$r \leq \min\{n,m\}$ is the rank of $N$, and $\kappa_i >0$, $i=1,\ldots,r$. Then, $G(s)$ can be factorized as $G(s)= V\,\hat{G}(s)\, (V^{\dag}S)$, where $\hat{G}(s)$ has the form
\[ \hat{G}(s)= I-\hat{N}\,\Big(sI + \imath\hat{M} + \frac{1}{2}\hat{N}^{\dag}\hat{N}\Big)^{-1} \, \hat{N}^{\dag}, \]
with $\hat{M}=W^{\dag}M W$. The first and last factors in this factorization of $G(s)$ are unitary transformations of the output and the input, respectively, of the transfer function $\hat{G}(s)$ in the middle factor, and can be realized by multi-beam splitters. The transfer function $\hat{G}(s)$ is that of a passive LQSS with scattering matrix $I$, coupling matrix $\hat{N}$, and Hamiltonian matrix $\hat{M}=W^{\dag}M W$. Moreover, $\hat{G}(s)$ can be realized by the following feedback network of $(n-r)$ 1-port and $r$ 2-port cavities:
\begin{eqnarray}
da &=& \Big(-\imath D - \frac{1}{2} \tilde{N}^{\dag} \tilde{N} - \frac{1}{2} \hat{N}^{\dag} \hat{N}\Big) a \,dt \nonumber \\
&-& \tilde{N}^{\dag} d\mathcal{U}_{int} - \hat{N}^{\dag} d\mathcal{U}, \nonumber \\
d\mathcal{Y} &=& \hat{N} a\, dt + d\mathcal{U}, \nonumber \\
d\mathcal{Y}_{int} &=& \tilde{N} a\, dt + d\mathcal{U}_{int}, \nonumber \\
d\mathcal{U}_{int} &=& R\, d\mathcal{Y}_{int}. \label{Feedback network realization, passive case eqn}
 \end{eqnarray}
Here, $D = \diag(\Delta_1,\ldots,\Delta_n)$, and $\tilde{N} = \diag(\sqrt{\tilde{\kappa}_1},\ldots,\sqrt{\tilde{\kappa}_n})$, where $\Delta_i \in \mathbb{R}$, and  $\tilde{\kappa}_i>0$, are the cavity detuning and the coupling coefficient of the interconnection port, respectively, of the $i$-th cavity, which can be chosen arbitrarily. The $m$-dimensional vectors $\mathcal{U}$, and $\mathcal{Y}$, contain the inputs/outputs of the system ports, and the $n$-dimensional vectors $\mathcal{U}_{int}$, and $\mathcal{Y}_{int}$, the inputs/outputs of the interconnection ports. Finally, the unitary interconnection matrix (feedback gain) $R$ is determined through the relations
\begin{eqnarray}
X &=& 2\imath \tilde{N}^{-\dag} (\hat{M}-D) \tilde{N}^{-1}, \label{solution for X, passive case} \\
R &=& (X-I)(X+I)^{-1}. \square \label{inverse Cayley transform}
\end{eqnarray}
\end{theorem}
From the fact that $D$, $\hat{N}$, and $\tilde{N}$ are diagonal, all diagonal elements of $\tilde{N}$ are non-zero, and only $r$ diagonal elements of $\hat{N}$ are non-zero, we see that (\ref{Feedback network realization, passive case eqn}) describes a collection of cavities, all of which have one interconnection port, but only $r$ have system ports. Hence, the feedback network consists of $(n-r)$ 1-port and $r$ 2-port cavities. Figure \ref{Passive_Network_3} provides a graphical representation of the realization method of Theorem \ref{Cascade realization, passive case thm}.

\noindent\textbf{Proof:} It suffices to prove that $\hat{G}(s)$ is the transfer function of the system described by (\ref{Feedback network realization, passive case eqn}). To this end, we combine the last two equations in (\ref{Feedback network realization, passive case eqn}) to obtain the relation $d\mathcal{U}_{int} = (I-R)^{-1} R\, \tilde{N} a\, dt$. At this point we introduce a variant of the Cayley transform for unitary matrices without unit eigenvalues \cite{golvan96}, namely
\begin{equation}\label{Cayley transform}
X=(I-R)^{-1}(I+R).
\end{equation}
The unitarity of $R$ implies that $X$ is skew-Hermitian. We can also solve uniquely for $R$ in terms of $X$ with the following result:
\[R=(X-I)(X+I)^{-1},\]
where $R$ is defined for all skew-Hermitian matrices $X$, and can be seen to be unitary due to the skew-Hermitian nature of $X$. It is easy to see that $(I-R)^{-1}R= - \frac{1}{2} I + \frac{1}{2}X$. Using the relation between $d\mathcal{U}_{int}$ and $a$, and the definition of $X$, the equations for the network take the following form:
\begin{eqnarray}
da &=& \Big(-\imath D - \frac{1}{2} \tilde{N}^{\dag} X \tilde{N} - \frac{1}{2} \hat{N}^{\dag} \hat{N}\Big) a \,dt - \hat{N}^{\dag} d\mathcal{U} ,  \nonumber \\
d\mathcal{Y} &=& \hat{N} a\, dt + d\mathcal{U}.  \label{Interconnected cavities model, passive case}
\end{eqnarray}
These equations describe a passive linear quantum stochastic system with Hamiltonian matrix $\hat{M}$ given by the expression
\begin{equation}\label{Network Hamiltonian, passive case}
\hat{M}=D - \frac{\imath}{2} \tilde{N}^{\dag} X \tilde{N}.
\end{equation}
Given any values for the cavity parameters $\Delta_i \in \mathbb{R}$ and $\tilde{\kappa}_i >0$, and any desired Hamiltonian matrix $\hat{M}=W^{\dag}M W$, we may determine the unique $X$ (and hence the unique $R$) that achieves this $\hat{M}$ by the expression
\[X=2\imath \tilde{N}^{-\dag} (\hat{M}-D) \tilde{N}^{-1}.\blacksquare \]

\begin{figure}[!h]
\begin{center}
\scalebox{.25}{\includegraphics{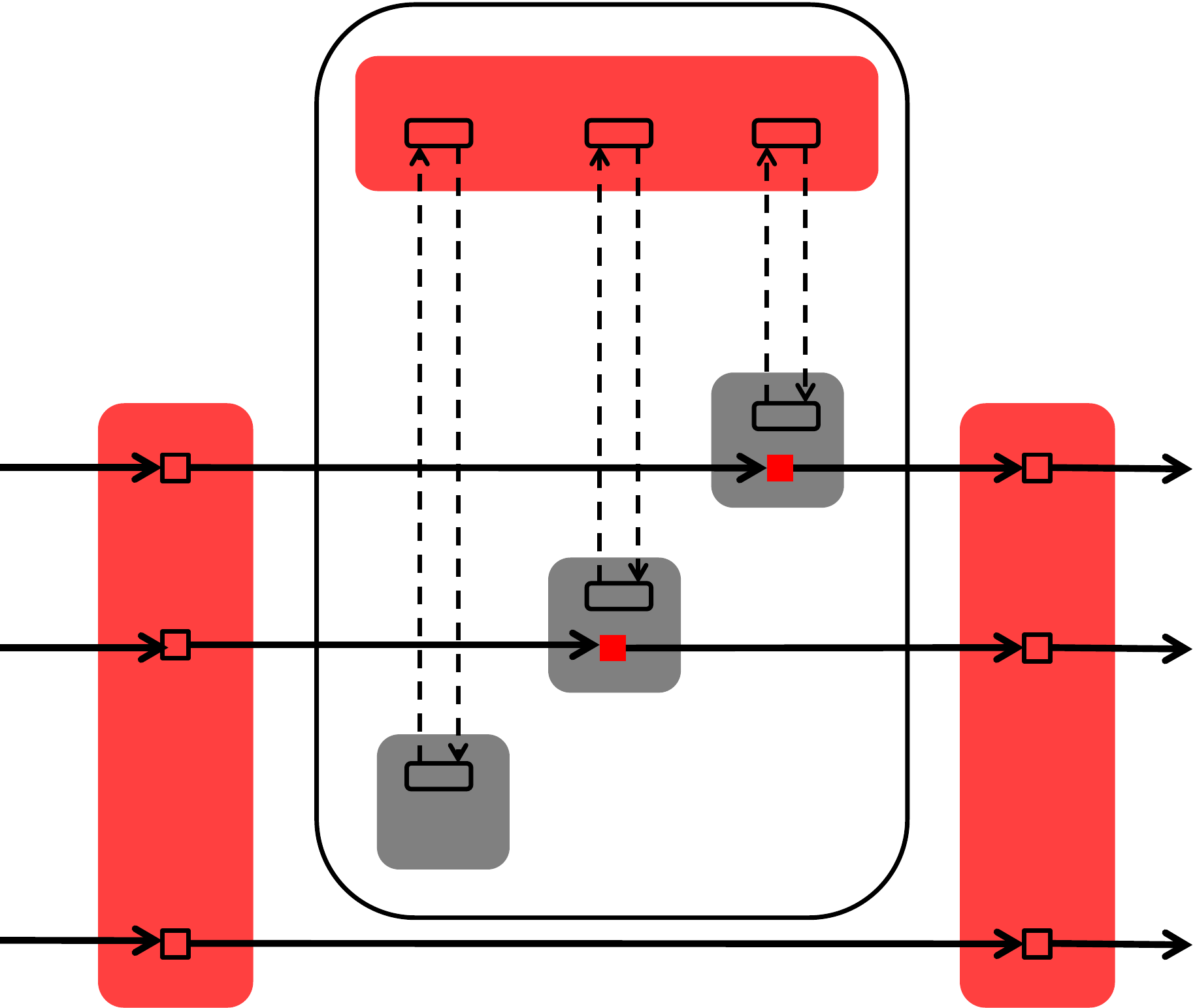}} \caption{Graphical representation of the proposed implementation of the passive transfer function in Example \ref{example 2}.} \label{Passive_Network_4}
\end{center}
\end{figure}

Similarly to the cascade realization, there is non-uniqueness associated with the ordering of the singular values of $N$ on the diagonal of $\hat{N}$. However, there is additional non-uniqueness due to a continuum of choices for the values of $\Delta_i$ and $\tilde{\kappa}_i$, $i=1,\ldots,n$. We demonstrate this method with an illustrative example.
\begin{example}\label{example 2}
For the system of Example \ref{example 1}, we have that the SVD of $N$ is given by $N=V\hat{N}W^{\dag}$, with
\begin{eqnarray*}
V &=& \left(\begin{array}{rrr}
-0.2987 & 0.4941 & -0.8165\\
-0.3065 & -0.8599 & -0.4082\\
-0.9038 & 0.1283 & 0.4082\\
\end{array}\right),\\
W &=& \left(\begin{array}{rrr}
-0.3093 & 0.2717 & -0.9113\\
-0.4409 & 0.8081 & 0.3906\\
-0.8426 & -0.5226 & 0.1302\\
\end{array}\right),\ \mathrm{and,} \\
\hat{N} &=& \diag( 6.8092, 2.7632, 0).
\end{eqnarray*}
The Hamiltonian of the reduced system is given by
\begin{equation*}
\hat{M}=W^{\dag}M W = \left(\begin{array}{rrr}
3.1315 & 0.0370 & -0.7200\\
0.0370 & 4.4278 & -2.2169\\
-0.7200 & -2.2169 & 4.4407\\
\end{array}\right).
\end{equation*}
Letting $D=0_{3 \times 3}$ and $\tilde{N}=I_3$, equation (\ref{solution for X, passive case}) produces the following $X$:
\begin{equation*}
X= \imath \left(\begin{array}{rrr}
6.2631 &  0.0740 & -1.4400 \\
0.0740 &  8.8556 & -4.4337 \\
-1.4400 & -4.4337 &  8.8814 \\
\end{array}\right),
\end{equation*}
from which we calculate the feedback gain matrix $R$ using equation (\ref{inverse Cayley transform}),
\begin{eqnarray*}
R&=&\left(\begin{array}{rrr}
0.9429 & -0.0145 & -0.0237 \\
-0.0145 & 0.9438 & -0.0467 \\
-0.0237 & -0.0467 & 0.9389 \\
\end{array} \right) \\
&+& \imath \left(\begin{array}{rrr}
0.3245 & 0.0276 & 0.0637 \\
0.0276 & 0.2918 & 0.1449 \\
0.0637 & 0.1449 & 0.3010 \\
\end{array} \right)
\end{eqnarray*}
Figure \ref{Passive_Network_4} provides a graphical representation of the proposed implementation of the transfer function for this example.$\square$
\end{example}

\section{Realizations of General Linear Quantum Stochastic Systems}
\label{Realizations of General Linear Quantum Stochastic Systems}

In this section, we extend the transfer function realization methods for passive LQSSs presented in Section \ref{Realizations of Passive Linear Quantum Stochastic Systems}, to general LQSSs.  These methods employed Schur's Unitary Triangularization theorem, and the Singular Value Decomposition \cite{horjoh85}, respectively. To extend the methods to the general case, we prove versions of these two classic matrix decompositions for doubled-up matrices in ($\mathbb{C}^{2k}$, $J_{2k}$).

\subsection{Cascade Realization}
\label{Cascade Realization, General Case}

We begin with the analog of Schur's Unitary Triangularization theorem for doubled-up matrices in Krein spaces. A version of this result for symplectic spaces has been derived in \cite{nurgripet16}. Here, we prove the Krein space version in a way that closely follows the proof of the classic result in \cite{horjoh85}.
\begin{lemma}\label{Bogoliubov Schur}
Let $A\in \mathbb{C}^{2n \times 2n}$ be a doubled-up matrix. Then, under Assumption I \footnote{See remark right after the proof.}, there is a Bogoliubov matrix $W^{2n \times 2n}$, such that $W^{\flat}AW=T=\bigl(\begin{smallmatrix} T_1 & T_2 \\ T_2^{\#} & T_1^{\#} \end{smallmatrix}\bigr)$, where $T_1$ is lower triangular and $T_2$ is strictly lower triangular. $\square$
\end{lemma}
\vspace*{.5em}
\textbf{Proof:} First, we prove certain facts about the eigenstructure of $A$. Let $\lambda$ be an eigenvalue of $A$ with corresponding eigenvector $v$, i.e. $Av=\lambda v$. We compute:
\begin{eqnarray*}
&& Av=\lambda v \Rightarrow A^{\#}v^{\#}=\lambda^{*} v^{\#} \Rightarrow (\Sigma A \Sigma) v^{\#}=\lambda^{*} v^{\#} \\
&\Rightarrow& A (\Sigma v^{\#})=\lambda^{*} (\Sigma v^{\#}).
\end{eqnarray*}
For nonreal $\lambda$, this implies that $\lambda^{*}$ is also an eigenvalue of $A$, with eigenvector $\Sigma v^{\#}$. For a real $\lambda$, there are two possibilities: $\Sigma v^{\#}$ is either linearly independent from $v$, or not. We show that, under the assumption that $v$ has non-zero $J$-norm, the second possibility cannot occur.
In fact, we shall prove that for any $u \in \mathbb{C}^{2n}$, $\Sigma u^{\#}$ is linearly independent from $u$, if $u^{\dag}J u \neq 0$. Indeed, if $\Sigma u^{\#}=\rho u$, for some $\rho \in \mathbb{C}^{*}$, then $u=\rho^{*}\Sigma u^{\#}$, and the two equations are compatible only if $|\rho|=1$. At this point, we introduce two identities which shall be useful in the following:
\begin{eqnarray}
(\Sigma w_1^{\#})^{\dag} J (\Sigma w_2^{\#}) &=& -(w_1^{\dag}J w_2)^*, \nonumber \\
(\Sigma w_1^{\#})^{\dag} J w_2 &=& -(w_1^{\dag}J \Sigma w_2^{\#})^*, \label{useful identities I}
\end{eqnarray}
for complex vectors $w_1, w_2$. Both identities can be proven using the simple relation $\Sigma J \Sigma=-J$. Immediate consequences of these are that, $w$ and $(\Sigma w^{\#})$ are $J$-orthogonal and have opposite $J$-norms, for any complex vector $w$, i.e.
\begin{eqnarray}
(\Sigma w^{\#})^{\dag} J (\Sigma w^{\#}) &=& -w^{\dag}J w,\ \mathrm{and} \nonumber \\
(\Sigma w^{\#})^{\dag} J w &=& 0. \label{useful identities II}
\end{eqnarray}
Then, we have that $-u^{\dag}J u = (\Sigma u^{\#})^{\dag} J (\Sigma u^{\#}) = |\rho|^2 u^{\dag}J u = u^{\dag}J u \Rightarrow u^{\dag}J u =0$, which is excluded by the assumption that $u$ has non-zero $J$-norm.

Let $(\lambda_n,v_n)$ be an eigenvalue/eigenvector pair of $A$. We shall assume that $v_n^{\dag}Jv_n \neq 0$, and in particular that $v_n^{\dag}Jv_n >0$. This guarantees that $v_n$ and $\Sigma v_n^{\#}$ are linearly independent.  If $v_n^{\dag}Jv_n <0$, we replace $v_n$ with $\Sigma v_n^{\#}$, and $\lambda_n$ with $\lambda_n^{*}$. Let $x_n=v_n/\sqrt{v_n^{\dag}Jv_n}$ be the $J$-normalized version of $v_n$. Then, $A\, x_n = \lambda_n x_n$, $x_n^{\dag}Jx_n=1$, and $(\Sigma x_n^{\#})^{\dag} J (\Sigma x_n^{\#})=-1$. We can always extend the set $\{x_n, \Sigma x_n^{\#} \}$ to a $J$-orthonormal basis $\{x_1,\ldots,x_n, \Sigma x_1^{\#},\ldots, \Sigma x_n^{\#} \}$ of $\mathbb{C}^{2n}$, where
\begin{eqnarray*}
x_i^{\dag} J x_j &=& \delta_{ij}, \\
(\Sigma x_i^{\#})^{\dag} J (\Sigma x_j^{\#}) &=& -\delta_{ij}, \\
(\Sigma x_i^{\#})^{\dag} J x_j &=& 0, \ i,j=1,\ldots,n.
\end{eqnarray*}
This is possible because the subspace of $\mathbb{C}^{2n}$ spanned by the set $\{x_n, \Sigma x_n^{\#} \}$ is non-degenerate (it has a basis of vectors with non-zero $J$-norms), hence its $J$-orthogonal complement in $\mathbb{C}^{2n}$ is non-degenerate \cite{gohlanrod83}, and thus it is spanned by a basis of vectors with non-zero $J$-norms. This implies that there exists a vector $y_{n-1}$, such that $y_{n-1}^{\dag} J x_n=0$, $y_{n-1}^{\dag} J \Sigma x_n^{\#}=0$, and $y_{n-1}^{\dag} J y_{n-1} \neq 0$.  Because of the identities (\ref{useful identities I}), one can show that $(\Sigma y_{n-1}^{\#})^{\dag} J x_n=0$, $(\Sigma y_{n-1}^{\#})^{\dag} J \Sigma x_n^{\#}=0$, and $(\Sigma y_{n-1}^{\#})^{\dag} J \Sigma y_{n-1} \neq 0$. If $y_{n-1}^{\dag} J y_{n-1}>0$, we let $x_{n-1}= y_{n-1}/\sqrt{y_{n-1}^{\dag}Jy_{n-1}}$, otherwise we let $x_{n-1}=\Sigma y_{n-1}^{\#}/\sqrt{|y_{n-1}^{\dag}Jy_{n-1}|}$. Then, $x_{n-1}$ satisfies $x_{n-1}^{\dag} J x_n=0$, $x_{n-1}^{\dag} J \Sigma x_n^{\#}=0$, and $x_{n-1}^{\dag} J x_{n-1} =1$. Similarly, $(\Sigma x_{n-1}^{\#})^{\dag} J x_n=0$, $(\Sigma x_{n-1}^{\#})^{\dag} J \Sigma x_n^{\#}=0$, and $(\Sigma x_{n-1}^{\#})^{\dag} J \Sigma x_{n-1} = -1$. Hence, we have constructed two more basis vectors. Continuing in this fashion, we complete the basis. Practically, this could be implemented by starting, for example, from the set $\{x_n, \Sigma x_n^{\#}, e_1,\ldots,e_{2n}\}$, where $e_1,\ldots,e_{2n}$ are the standard basis vectors of $\mathbb{C}^{2n}$, and applying the Gram-Schmidt procedure in Krein space ($\mathbb{C}^{2n}$, $J_{2n}$) \cite{gohlanrod83}.

Now, define $W^{(n)} = \big[\, [x_1 \,\ldots\, x_n] \ \Sigma [x_1 \,\ldots\, x_n]^{\#} \big]$. $W^{(n)}$ is Bogoliubov, and
\begin{eqnarray*}
A W^{(n)}=W^{(n)} \left(\begin{array}{cc|cc}
A_1^{(n-1)} & \mathbf{0} & A_2^{(n-1)} & \mathbf{0} \\
\star & \lambda_n & \star & 0 \\ \hline
A_2^{(n-1)\,\#} & \mathbf{0} & A_1^{(n-1)\,\#} & \mathbf{0} \\
\star & 0 & \star & \lambda_n^* \\
\end{array}\right).
\end{eqnarray*}
The matrix $A^{(n-1)}=\bigl(\begin{smallmatrix} A_1^{(n-1)} & A_2^{(n-1)} \\ A_2^{(n-1)\,\#} & A_1^{(n-1)\,\#}  \end{smallmatrix}\bigr)$ is obviously doubled-up. Repeating the above procedure, we can find a $2(n-1)\times 2(n-1)$ Bogoliubov matrix $\tilde{W}^{(n-1)}$ and a $\lambda_{n-1} \in \mathbb{C}$, such that
\begin{eqnarray*}
&&A^{(n-1)} \tilde{W}^{(n-1)} \\
&=&\tilde{W}^{(n-1)} \left(\begin{array}{cc|cc}
A_1^{(n-2)} & \mathbf{0} & A_2^{(n-2)} & \mathbf{0} \\
\star & \lambda_{n-1} & \star & 0 \\ \hline
A_2^{(n-2)\,\#} & \mathbf{0} & A_1^{(n-2)\,\#} & \mathbf{0} \\
\star & 0 & \star & \lambda_{n-1}^* \\
\end{array}\right).
\end{eqnarray*}
Now, define
\begin{eqnarray*}
W^{(n-1)}=\left(\begin{array}{cc|cc}
\tilde{W}_1^{(n-1)} & \mathbf{0} & \tilde{W}_2^{(n-1)} & \mathbf{0} \\
\mathbf{0} & 1 & \mathbf{0} & 0 \\ \hline
\tilde{W}_2^{(n-1)\,\#} & \mathbf{0} & \tilde{W}_1^{(n-1)\,\#} & \mathbf{0} \\
\mathbf{0} & 0 & \mathbf{0} & 1 \\
\end{array}\right).
\end{eqnarray*}
We see that $W^{(n-1)}$ is a $2n \times 2n$ Bogoliubov matrix, and that
\begin{eqnarray*}
&& A W^{(n)} W^{(n-1)}= W^{(n)} W^{(n-1)} \\[.25em]
&\times&\left(\begin{array}{ccc|ccc}
A_1^{(n-2)}& \mathbf{0} & \mathbf{0} & A_2^{(n-2)}& \mathbf{0} & \mathbf{0}  \\
\star & \lambda_{n-1} & 0 & \star & 0 & 0 \\
\star & \star & \lambda_n & \star & \star  & 0 \\ \hline
A_2^{(n-2)\,\#}& \mathbf{0} & \mathbf{0} & A_1^{(n-2)\,\#}& \mathbf{0} & \mathbf{0} \\
\star & 0 & 0 & \star & \lambda_{n-1}^* & 0 \\
\star & \star  & 0 & \star & \star & \lambda_n^* \\
\end{array}\right).
\end{eqnarray*}
Continuing in this fashion, we produce Bogoliubov matrices $W^{(n)}$, $W^{(n-1)}$, $\ldots W^{(2)}$, such that
\[ A W^{(n)} W^{(n-1)}\cdots W^{(2)} =  W^{(n)} W^{(n-1)}\cdots W^{(2)} T, \]
where $T$ has the structure announced in the statement of the lemma. Defining $W=W^{(n)} W^{(n-1)}\cdots W^{(2)}$ provides the desired decomposition, because $W$ is Bogoliubov, being the product of Bogoliubov matrices. For this algorithm to work, one must guarantee that at least one eigenvector of the matrices $A^{(n)}=A, A^{(n-1)},\ldots,A^{(2)}$ appearing in successive steps of the algorithm, has non-zero $J$-norm. $\blacksquare$

\begin{figure}[!h]
\begin{center}
\scalebox{.4}{\includegraphics{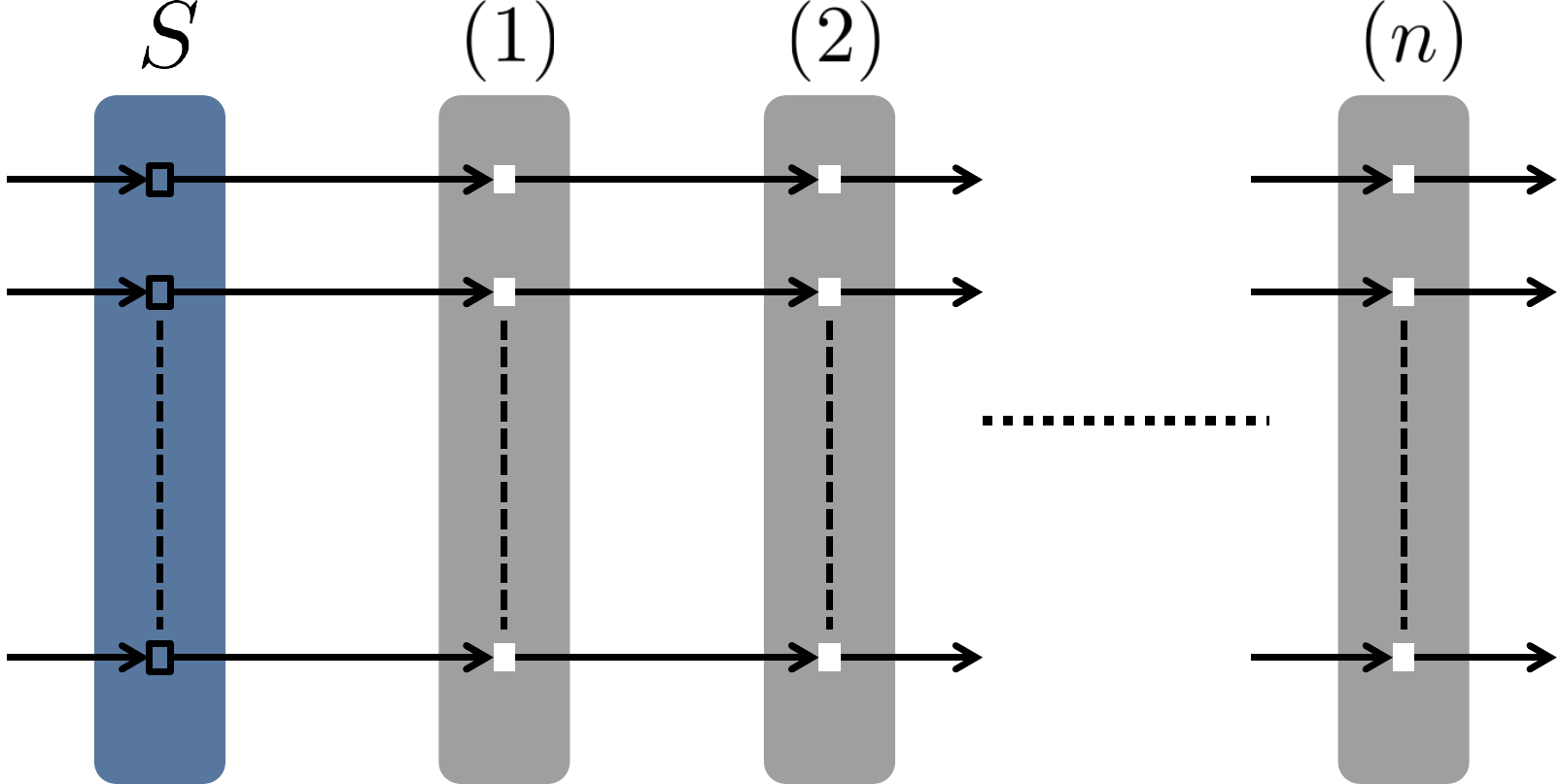}} \caption{Graphical representation of the cascade realization method for general LQSS transfer functions.} \label{General_Cascade}
\end{center}
\end{figure}

In order to emphasize it, we restate the sufficient condition for the algorithm in the proof of Lemma \ref{Bogoliubov Schur} to work:\\[.5em]
\emph{Assumption I}: At least one eigenvector of the matrices $A^{(n)}=A, A^{(n-1)},\ldots,A^{(2)}$ appearing in successive steps of the algorithm in the proof of Lemma \ref{Bogoliubov Schur}, must have non-zero $J$-norm.\\[.5em]
We should point out that, the factorization of Lemma \ref{Bogoliubov Schur}, is similar, but not identical, to the one obtained in \cite{nurgripet16}, restated in the context of Krein spaces and doubled-up matrices. A crucial difference is that, the factorization in this paper requires a strictly lower triangular $T_2$ matrix, while no such restriction is present in the approach of \cite{nurgripet16} in the real symplectic setting, restated in the context of Krein spaces and doubled-up matrices. A consequence of this is that, in Theorem \ref{Cascade realization, general case thm}, the dynamics of the annihilation operator $a_j$ of the $(j)$-th cavity does not depend on the dynamics of the corresponding creation operator $a_j^*$ of the same mode, and vice versa. Also, while the symplectic space version has been shown to hold for generic matrices, we have no such proof for Lemma \ref{Bogoliubov Schur}. Using Lemma \ref{Bogoliubov Schur}, we can extend the cascade realization of Subsection \ref{Cascade Realization, Passive Case} to general LQSSs:
\begin{theorem}\label{Cascade realization, general case thm}
Given a linear quantum stochastic system with Hamiltonian matrix $M^{2n \times 2n}$, coupling operator $N^{2m \times 2n}$, and generalized scattering matrix $S^{2m \times 2m}$, its transfer function can be realized by the following cascade of a multi-squeezer and $n$ $m$-port cavities:
\begin{eqnarray}
\check{\mathcal{Y}}_{(0)} &=& S\check{\mathcal{U}}, \nonumber \\
\mathcal{U}_{(1)} &=& \mathcal{Y}_{(0)}, \nonumber \\
da_1 &=& \Big[\!\!-\imath\Delta_{(1)} -\frac{1}{2}\big(N_{(1),1}^{\dag}N_{(1),1} -N_{(1),2}^{\top}N_{(1),2}^{\#}\big)\Big]  \nonumber \\
&& \times a_1\, dt -N_{(1),1}^{\dag}d\mathcal{U}_{(1)} + N_{(1),2}^{\top} d\mathcal{U}_{(1)}^{\#} \nonumber \\
d\mathcal{Y}_{(1)} &=& N_{(1),1} a_1 dt + N_{(1),2} a_1^{\#} dt +  d\mathcal{U}_{(1)}, \nonumber \\
\mathcal{U}_{(2)} &=& \mathcal{Y}_{(1)}, \nonumber \\
da_2 &=& \Big[\!\!-\imath\Delta_{(2)} -\frac{1}{2}\big(N_{(2),1}^{\dag}N_{(2),1} -N_{(2),2}^{\top}N_{(2),2}^{\#}\big)\Big]  \nonumber \\
&& \times a_2\, dt -N_{(2),1}^{\dag}d\mathcal{U}_{(2)} + N_{(2),2}^{\top} d\mathcal{U}_{(2)}^{\#} \nonumber \\
d\mathcal{Y}_{(2)} &=& N_{(2),1} a_2 dt + N_{(2),2} a_2^{\#} dt +  d\mathcal{U}_{(2)}, \nonumber \\
&\vdots& \nonumber \\
\mathcal{U}_{(n)} &=& \mathcal{Y}_{n-1}, \nonumber \\
da_n &=& \Big[\!\!-\imath\Delta_{(n)} -\frac{1}{2}\big(N_{(n),1}^{\dag}N_{(n),1} -N_{(n),2}^{\top}N_{(n),2}^{\#}\big)\Big]  \nonumber \\
&& \times a_n\, dt -N_{(n),1}^{\dag}d\mathcal{U}_{(n)} + N_{(n),2}^{\top} d\mathcal{U}_{(n)}^{\#} \nonumber \\
d\mathcal{Y}_{(n)} &=& N_{(n),1} a_n dt + N_{(n),2} a_n^{\#} dt +  d\mathcal{U}_{(n)}, \nonumber \\
\mathcal{Y} &=& \mathcal{Y}_{(n)}. \label{Cascade realization, general case eqn}
\end{eqnarray}
The cavity parameters $\Delta_{(i)} \in \mathbb{R}$, and $N_{(i)} \in \mathbb{C}^{2m \times 2}$, $i=1,\ldots,n$, are determined as follows: Define $F=-\imath JM -\frac{1}{2}N^{\flat}N$, and let $V$ a Bogoliubov matrix such that $VFV^{\flat}=VFV^{-1}$ has the structure described in Lemma \ref{Bogoliubov Schur}. Then, $\Delta_{(i)}= -\Im(VFV^{\flat})_{1,ii}$, $[N_{(1),1} N_{(2),1} \ldots N_{(n),1}]=(NV^{\flat})_1$, and $[N_{(1),2} N_{(2),2} \ldots N_{(n),2}]=(NV^{\flat})_2$, where the convention that, for a doubled-up matrix $X$, $X_1$ and $X_2$ will denote its upper-left and upper-right blocks (see Subsection \ref{Notation and terminology}), was used. This realization is possible if $F$ satisfies Assumption I. $\square$
\end{theorem}
Figure \ref{General_Cascade} provides a graphical representation of the cascade realization method of Theorem \ref{Cascade realization, general case thm}.

\noindent\textbf{Proof:} It is a straightforward calculation to show that the cascade system (\ref{Cascade realization, general case eqn}) can be put in the following form:
\begin{eqnarray*}
d\check{a} &=& \hat{F}\,\check{a}\, dt -\hat{N}^{\dag}S\, d\check{\mathcal{U}}, \\
d\check{\mathcal{Y}} &=& \hat{N} \check{a}\, dt + S\, d\check{\mathcal{U}},
\end{eqnarray*}
where $a=(a_1,\ldots,a_n)^{\top}$. $\hat{N}$ is doubled-up, with
\begin{eqnarray*}
\hat{N}_1 &=& [N_{(1),1} N_{(2),1} \ldots N_{(n),1}],\ \text{and} \\
\hat{N}_2 &=& [N_{(1),2} N_{(2),2} \ldots N_{(n),2}].
\end{eqnarray*}
$\hat{F}$ is doubled-up, with $\hat{F}_1$ lower triangular, and $\hat{F}_2$ strictly lower triangular. Their matrix elements are given by
\begin{eqnarray*}
\hat{F}_{1,ii}&=& -\imath\Delta_{(i)} -\frac{1}{2}\big(N_{(i),1}^{\dag}N_{(i),1} - N_{(i),2}^{\top}N_{(i),2}^{\#}\big), \\
\hat{F}_{1,ij}&=&- N_{(i),1}^{\dag}N_{(j),1} + N_{(i),2}^{\top}N_{(j),2}^{\#}, j < i, \\
\hat{F}_{2,ij}&=&- N_{(i),1}^{\dag}N_{(j),2} + N_{(i),2}^{\top}N_{(j),1}^{\#}, j < i, \\
\end{eqnarray*}
for $i,j=1,\ldots,n$.

Given a LQSS with parameters $(M,N,S)$, let $F=-\imath JM -\frac{1}{2}N^{\flat}N$. From Lemma \ref{Bogoliubov Schur}, for generic $F$ there exists a Bogoliubov matrix $V$ such that $\hat{F}=VFV^{\flat}=VFV^{-1}$ has a lower triangular $\hat{F}_1$, and a strictly lower triangular $\hat{F}_2$. Using $V$ as a state transformation, we obtain a realization of the system dynamics in cascade form. The parameters of the cascade realization are given in terms of the original parameters by $\hat{N}=NV^{-1}= NV^{\flat}$, and $\Delta_{(i)}= -\Im\hat{F}_{1,ii}, i=1,\ldots,n$. Since the transfer function of a linear system is independent of its particular realization, it follows that the transfer function of a LQSS can always be realized by the cascade form given in the theorem.$\blacksquare$

As in the passive case, the different choices of eigenvalues in every step of the algorithm described in Lemma \ref{Bogoliubov Schur} lead to different matrices $V$ and $\hat{F}$, and hence, to different realizations of the LQSS. We demonstrate this method with an illustrative example.
\begin{example}\label{example 3}
Consider the 2-mode, 2-input linear quantum stochastic system with the following parameters:
\begin{eqnarray*}
M = \left(\begin{array}{rrrr}
2  &   1  &   0   &  -1\\
1  &   2  &  -1   &  0\\
0  &  -1  &   2   &  1\\
-1 &   0  &   1   &  2\\
\end{array}\right), \\
N = \left(\begin{array}{rrrr}
0&1&2&0\\
-1&2&1&-1\\
2&0&0&1\\
1&-1&-1&2\\
\end{array}\right),
\end{eqnarray*}
and $S=I_4$. We compute $F=-\imath JM -\frac{1}{2}N^{\flat}N$ to be equal to
\begin{eqnarray*}
F = \left(\begin{array}{rrrr}
2-2i&0.5-1i&0&1.5+1i\\
0.5-1i&-2-2i&-1.5+1i&0\\
0&1.5-1i&2+2i&0.5+1i\\
-1.5-1i&0&0.5+1i&-2+2i\\
\end{array}\right).
\end{eqnarray*}
Using the algorithm of Lemma \ref{Bogoliubov Schur}, we compute a Bogoliubov matrix $V$ and a doubled-up $\hat{F}$ with $\hat{F}_1$ lower triangular, and $\hat{F}_2$ strictly lower triangular, such that $\hat{F}=VFV^{\flat}=VFV^{-1}$:
\begin{eqnarray*}
V &=&\left(\begin{array}{rrrr}
-0.1229&1.0111&0.0643&-0.0352\\
1.0200&0.1032&-0.1333&0.0529\\
0.0643&-0.0352&-0.1229&1.0111\\
-0.1333&0.0529&1.0200&0.1032\\
\end{array}\right) \\
&+& \imath\left(\begin{array}{rrrr}
0.2393&0.0226&-0.2991&-0.0219\\
0&0.2176&0.0392&-0.2764\\
0.2991&0.0219&-0.2393&-0.0226\\
-0.0392&0.2764&0&-0.2176\\
\end{array}\right), \\
\hat{F} &=& \left(\begin{array}{rrrr}
-2.0305&0&0&0\\
-0.1106&2.0305&2.4537&0\\
0&0&-2.0305&0\\
2.4537&0&-0.1106&2.0305\\
\end{array}\right) \\
&+& \imath \left(\begin{array}{rrrr}
-2.2667&0&0&0\\
-1.5909&-2.6660&2.2391&0\\
0&0&2.2667&0\\
-2.2391&-0&1.5909&2.6660\\
\end{array}\right).
\end{eqnarray*}
The parameters $M_{(1)}$,$M_{(2)} \in \mathbb{R}$, and $N_{(1)}$,$N_{(2)} \in \mathbb{C}^2$ are given by $(M_{(1)},M_{(2)})=-\Im(\hat{F}_{11},\hat{F}_{22})=(2.2667,2.6660)$, and
\begin{eqnarray*}
&& \hat{N}=NV^{\flat}=NV^{-1} \\
&=& \left(\begin{array}{rrrr}
0.8826&0.3697&-0.2106&1.9872\\
2.0457&-0.6276&-0.9994&0.6779\\
-0.2106&1.9872&0.8826&0.3697\\
-0.9994&0.6779&2.0457&-0.6276\\
\end{array}\right) \\
&+& \imath \left(\begin{array}{rrrr}
-0.6207&-0.1391&0.5005&0.2764\\
-0.0830&-0.1196&-0.0385&0.3744\\
-0.5005&-0.2764&0.6207&0.1391\\
0.0385&-0.3744&0.0830&0.1196\\
\end{array}\right). \square
\end{eqnarray*}
\end{example}

\subsection{Realization Using Static Networks for Input/Output Processing and Feedback}
\label{Realization Using Static Networks for Input/Output Processing and Feedback, General Case}

\begin{figure}[!h]
\begin{center}
\scalebox{.275}{\includegraphics{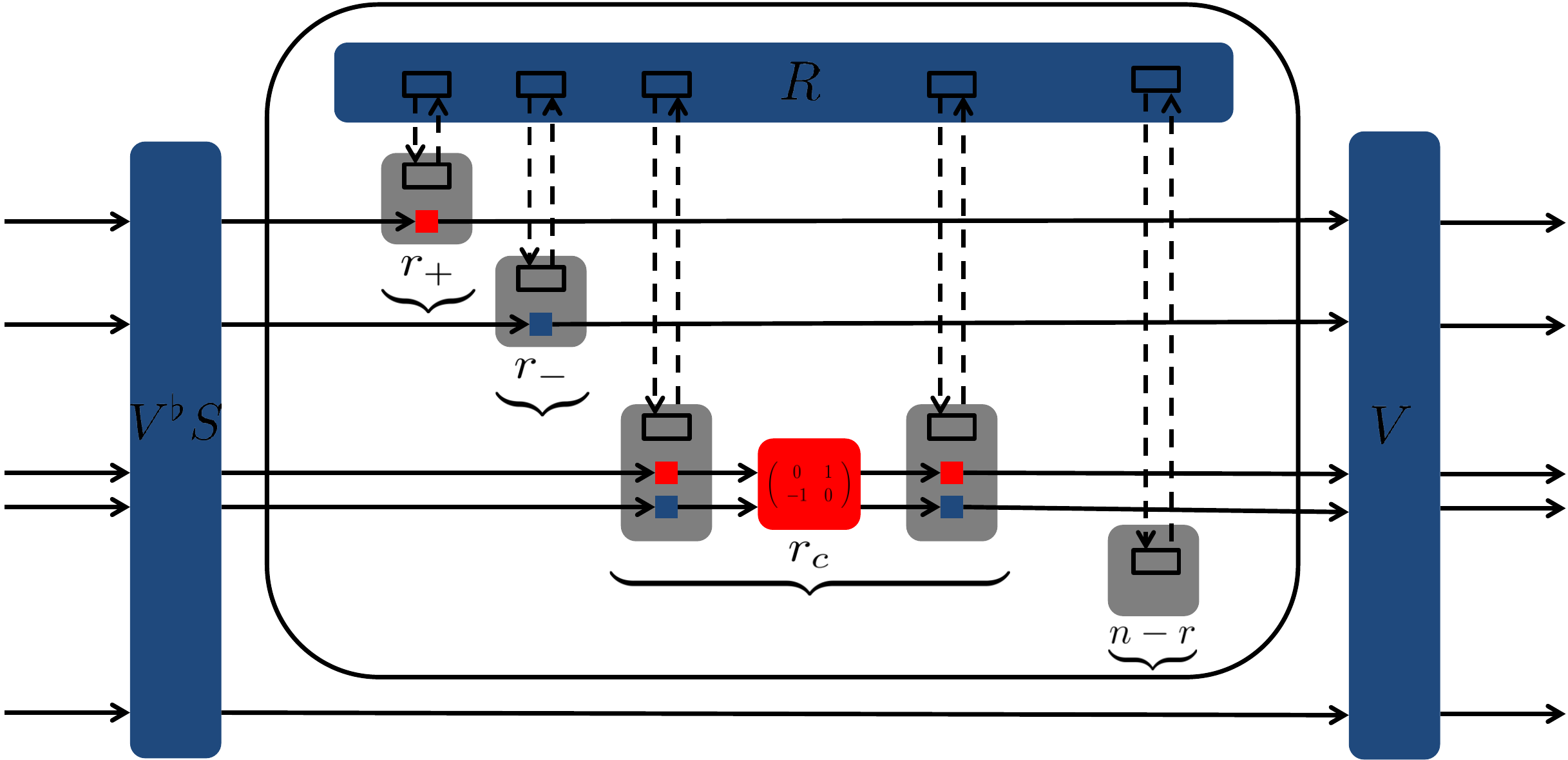}} \caption{A graphical representation of the realization of the transfer function of a general LQSS proposed in Theorem \ref{Feedback network realization, general case thm}. Each cavity is representative of all cavities of its type needed to implement the transfer function.} \label{Active_Network_1}
\end{center}
\end{figure}

\begin{figure}[!h]
\begin{center}
\scalebox{.275}{\includegraphics{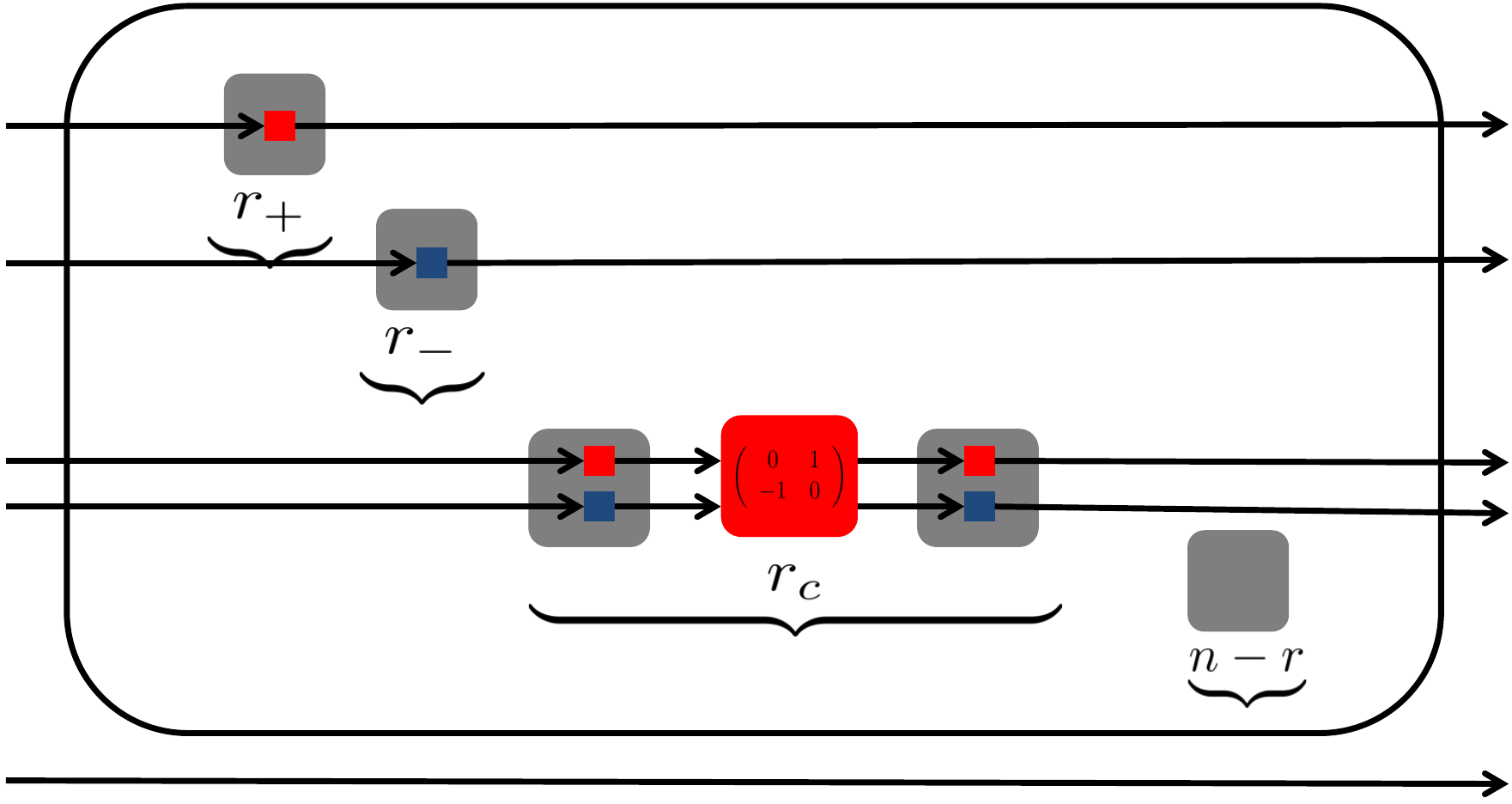}} \caption{A graphical representation of the LQSS (\ref{Interconnected cavities model, general case}) Each cavity is representative of all cavities of its type needed to implement the LQSS.} \label{Active_Network_1_5}
\end{center}
\end{figure}

To extend the corresponding realization method to the general case, we need an SVD-like decomposition for doubled-up matrices in Krein spaces \cite{gripet15}:
\begin{lemma} \label{Bogoliubov SVD}
Let $N \in \mathbb{C}^{2m \times 2n}$ be a doubled-up matrix, and let $\mathcal{N} = N^{\flat}N$. We assume that all the eigenvalues of $\mathcal{N}$ are semisimple, and that $\ker \mathcal{N}=\ker N$. Let $\lambda_i^+ >0, i=1,\ldots,r_+$, $\lambda_i^- <0, i=1,\ldots,r_-$, and $\lambda_i^c$ with $\Im\lambda_i>0$, $i=1\ldots,r_c$, be the eigenvalues of $\mathcal{N}$ that are, respectively, positive, negative, and non-real with positive imaginary part. Then, there exist Bogoliubov matrices $V^{2m \times 2m}$, $W^{2n \times 2n}$, and a doubled-up matrix $\hat{N} \in \mathbb{C}^{2m \times 2n}$, such that $N = V \, \hat{N} \, W^{\flat}$, where $\hat{N}_{1}= \bigl(\begin{smallmatrix} \bar{N}_{1} & \mathbf{0} \\ \mathbf{0} & \mathbf{0} \end{smallmatrix}\bigr)$, $\hat{N}_{2}= \bigl(\begin{smallmatrix} \bar{N}_{2} & \mathbf{0} \\ \mathbf{0} & \mathbf{0} \end{smallmatrix}\bigr)$, and
\begin{eqnarray*}
\bar{N}_1 &=& \diag(\sqrt{\lambda_1^+},\ldots,\sqrt{\lambda_{r_+}^+},\underbrace{0,\ldots,0}_{r_-},\alpha_1 I_2,\ldots,\alpha_{r_c} I_2), \\
\bar{N}_2 &=& \diag(\underbrace{0,\ldots,0}_{r_+},\sqrt{|\lambda_1^-|},\ldots,\sqrt{|\lambda_{r_-}^-|},-\beta_1\sigma_2,\ldots, \\
&& -\beta_{r_c}\sigma_2).
\end{eqnarray*}
The parameters $\alpha_i$ and $\beta_i$ are determined in terms of $\lambda_i^c$, as follows:
\[ \alpha_i = \sqrt{\frac{|\lambda_i^c|+\Re \lambda_i^c}{2}}, \ \ \beta_i = \frac{\Im \lambda_i^c}{\sqrt{2\big(|\lambda_i^c|+\Re \lambda_i^c\big)}}.\, \square \]
\end{lemma}
The proof of the lemma can be found in \cite{gripet15}. Using Lemma \ref{Bogoliubov SVD}, we may extend the feedback network realization of Subsection \ref{Realization Using Static Networks for Input/Output Processing and Feedback, Passive Case} to general LQSSs:
\begin{theorem}\label{Feedback network realization, general case thm}
Given a linear quantum stochastic system with Hamiltonian matrix $M^{2n \times 2n}$, coupling operator $N^{2m \times 2n}$, and generalized scattering matrix $S^{2m \times 2m}$, let
\[ G(s)=\Big[ I-N\, \Big(sI+\imath JM+\frac{1}{2}\, N^{\flat}N\Big)^{-1}N^{\flat}\Big] S \]
be its transfer function. If $N$ satisfies the assumptions of Lemma \ref{Bogoliubov SVD}, let $N=V\hat{N}W^{\flat}$ be the decomposition of the coupling matrix $N$ according to Lemma \ref{Bogoliubov SVD}. Then, $G(s)$ can be factorized as $G(s)= V\,\hat{G}(s)\, (V^{\flat}S)$, where $\hat{G}(s)$ has the form
\[ \hat{G}(s)= I-\hat{N}\,\Big(sI + \imath J\hat{M} + \frac{1}{2}\hat{N}^{\flat}\hat{N}\Big)^{-1} \, \hat{N}^{\flat}, \]
with $\hat{M}=W^{\dag}M W$. The first and last factors in this factorization of $G(s)$ are Bogoliubov transformations of the output and the input, respectively, of the transfer function $\hat{G}(s)$ in the middle factor, and can be realized by multi-squeezers. The transfer function $\hat{G}(s)$ is that of a LQSS with scattering matrix $I$, coupling matrix $\hat{N}$, and Hamiltonian matrix $\hat{M}$. Moreover, $\hat{G}(s)$ can be realized by the following feedback network of $(n-r)$ 1-port, $(r_+ + r_-)$ 2-port, and $2r_c$ 3-port cavities, where $r=(r_+ + r_- +2r_c)$:
\begin{eqnarray}
d\check{a} &=& [-\imath J\bar{M} -\frac{1}{2}\tilde{N}^{\flat}\tilde{N} - \frac{1}{2} \hat{N}^{\flat}\hat{ N}]\, \check{a} dt
- \tilde{N}^{\flat} d\check{\mathcal{U}}_{int} \nonumber \\
&-& \hat{N}^{\flat} d\check{\mathcal{U}}, \nonumber \\
d\check{\mathcal{Y}} &=& \hat{N} \check{a} dt + d\check{\mathcal{U}}, \nonumber \\
d\check{\mathcal{Y}}_{int} &=& \tilde{N} \check{a} dt + d\check{\mathcal{U}}_{int} , \nonumber \\
d\check{\mathcal{U}}_{int} &=& R\, d\check{\mathcal{Y}}_{int}. \label{Feedback network realization, general case eqn}
 \end{eqnarray}
Here, $\bar{M} = \diag(D,D) +E+E^{\top}$, where
\begin{eqnarray*}
D &=& \diag(\Delta_1^+,\ldots,\Delta_{r_+}^+,\Delta_1^-,\ldots,\Delta_{r_-}^-, \\
&& \Delta_1^c,\Delta_1^c, \ldots,\Delta_{r_c}^c,\Delta_{r_c}^c,\Delta_1^0,\ldots,\Delta_{n-r}^0),
\end{eqnarray*}
and $E^{2n \times 2n}$ has all zero elements except for
\begin{eqnarray*}
E_{r_+ + r_- +2i-1,n+r_+ + r_- +2i} &=& -\frac{\Im\lambda_i^c}{2}, \\
E_{r_+ + r_- +2i,n+r_+ + r_- +2i-1} &=& -\frac{\Im\lambda_i^c}{2},
\end{eqnarray*}
for $i=1,\ldots,r_c$. $\tilde{N}= \diag(\sqrt{\tilde{\kappa}_1},\ldots,\sqrt{\tilde{\kappa}_n},\sqrt{\tilde{\kappa}_1},\ldots$, $\sqrt{\tilde{\kappa}_n})$. The $\Delta$'s, and  $\tilde{\kappa}$'s, are cavity detunings and coupling coefficients of the (passive) interconnection ports, respectively, of individual cavities, and can be chosen arbitrarily. The $m$-dimensional vectors $\mathcal{U}$, and $\mathcal{Y}$, contain the inputs/outputs of the system ports, and the $n$-dimensional vectors $\mathcal{U}_{int}$, and $\mathcal{Y}_{int}$, the inputs/outputs of the interconnection ports. Finally, the Bogoliubov interconnection matrix (feedback gain) $R$ is determined through the relations
\begin{eqnarray}
X &=& 2\imath (\tilde{N}^{\flat})^{-1} (J\hat{M}-J\bar{M})\, \tilde{N}^{-1}, \label{solution for X, general case} \\
R &=& (X-I)(X+I)^{-1}.\,\square \nonumber
\end{eqnarray}
\end{theorem}
Figure \ref{Active_Network_1} is a graphical representation of the realization method proposed in Theorem \ref{Feedback network realization, general case thm}. \\
%
\textbf{Proof:} The proof consists of two parts. First, we show that the LQSS
\begin{eqnarray}
d\check{a} &=& [-\imath J\bar{M} - \frac{1}{2} \hat{N}^{\flat}\hat{ N}]\, \check{a} dt - \hat{N}^{\flat} d\check{\mathcal{U}}, \nonumber \\
d\check{\mathcal{Y}} &=& \hat{N} \check{a} dt + d\check{\mathcal{U}}, \label{Interconnected cavities model, general case}
\end{eqnarray}
represents a collection of cavities as announced in the theorem, see Figure \ref{Active_Network_1_5}.
If we look at the structure of $\hat{N}$ and $\bar{M}$, we conclude the following:
\begin{enumerate}
  \item Part of the system consists of $r_+$ independent passive cavities with Hamiltonian matrices $\diag(\Delta_i^+,\Delta_i^+)$, and coupling matrices $\diag(\sqrt{\lambda_i^+},\sqrt{\lambda_i^+})$, $i=1,\ldots,r_+$, corresponding to the positive eigenvalues of $\mathcal{N}$.
  \item Part of the system consists of $r_-$ independent purely active cavities with Hamiltonian matrices $\diag(\Delta_i^-,\Delta_i^-)$, and coupling matrices $\bigl(\begin{smallmatrix} 0 & \sqrt{|\lambda_i^-|} \\ \sqrt{|\lambda_i^-|} & 0 \end{smallmatrix} \bigr)$, $i=1,\ldots,r_-$, corresponding to the negative eigenvalues of $\mathcal{N}$.
  \item Part of the system consists of $r_c$ independent LQSSs with two modes and two inputs/outputs, with Hamiltonian matrices
  \[ \left(\begin{array}{cccc}
\Delta_i^c & 0 & 0 & -\Im\lambda_i^c/2 \\
0 & \Delta_i^c & -\Im\lambda_i^c/2 & 0 \\
0 & -\Im\lambda_i^c/2 & \Delta_i^c & 0 \\
-\Im\lambda_i^c/2 & 0 & 0 & \Delta_i^c \\
\end{array}\right),  \]
  and coupling matrices
  \[ \left(\begin{array}{rrrr}
\alpha_i & 0 & 0 & \imath \beta_i \\
0 & \alpha_i & -\imath\beta_i & 0 \\
0 & -\imath\beta_i & \alpha_i & 0 \\
\imath\beta_i & 0 & 0 & \alpha_i  \\ \end{array}\right), \]
  for $i=1,\ldots,r_c$, corresponding to the nonreal eigenvalues of $\mathcal{N}$. One can realize such a LQSS as a cascade connection of two identical 2-port cavities and a beam-splitter, as in Figure \ref{Active_Network_1_5}. Each cavity has two ports, one passive with coupling coefficient $\alpha_i^2$, and one purely active with coupling coefficient $\beta_i^2$. Its coupling matrix is given by
\[ \left(\begin{array}{rr}
0 & \imath\beta_i \\
\alpha_i & 0 \\
-\imath\beta_i & 0 \\
0 & \alpha_i \\
\end{array}\right), \]
and its Hamiltonian matrix by $\diag(\Delta_i^c,\Delta_i^c)$, for $i=1,\ldots,r_c$. The beam splitter implements the unitary transformation $\bigl(\begin{smallmatrix} 0 & 1 \\ -1 & 0 \end{smallmatrix} \bigr)$.
  \item The rest of the system consists of $n-r$ unconnected cavities with Hamiltonian matrices $\diag(\Delta_i^0,\Delta_i^0)$, and coupling matrices equal to zero.
\end{enumerate}
Note that the introduction of the interconnection adds an extra passive port per cavity, see Figure \ref{Active_Network_1}. To complete the proof, it suffices to prove that $\hat{G}(s)$ is the transfer function of the feedback network  described by (\ref{Feedback network realization, general case eqn}). To this end, we combine the last two equations in (\ref{Feedback network realization, general case eqn}) to obtain the relation $d\check{\mathcal{U}}_{int} = (I-R)^{-1} R\, \tilde{N} \check{a}\, dt$. Now we introduce the variant of the Cayley transform (\ref{Cayley transform}), $X=(I-R)^{-1}(I+R)$, for Bogoliubov matrices without unit eigenvalues. It is straightforward to verify that $X$ is doubled-up and $\flat$-skew-Hermitian ($X^{\flat}=-X$) if and only if $R$ is Bogoliubov. The unique solution for $R$ in terms of $X$ is given by (\ref{inverse Cayley transform}), $R=(X-I)(X+I)^{-1}$, where $R$ is defined for all $\flat$-skew-Hermitian matrices $X$. Using the identity $(I-R)^{-1}R= - \frac{1}{2} I + \frac{1}{2}X$, the relation between $d\check{\mathcal{U}}_{int}$ and $\check{a}$, and the definition of $X$, the equations for the feedback network take the following form:
\begin{eqnarray*}
d\check{a} &=& [-\imath J \bar{M} -\frac{1}{2}\tilde{N}^{\flat} X \tilde{N} - \frac{1}{2} \hat{N}^{\flat}\hat{ N}]\, \check{a} dt - \hat{N}^{\flat} d\check{\mathcal{U}},  \\
d\check{\mathcal{Y}} &=& \hat{N} \check{a} dt + d\check{\mathcal{U}}.
 \end{eqnarray*}
These equations describe a LQSS with Hamiltonian matrix $\hat{M}$ given by the expression
\begin{equation}\label{Network Hamiltonian, general case}
J\hat{M}= J\bar{M}-\frac{\imath}{2} (\tilde{N}^{\flat} X \tilde{N}).
\end{equation}
Given any values for the cavity parameters $\Delta$ and $\tilde{\kappa}$, and any desired Hamiltonian matrix $\hat{M}=W^{\dag}M W$, we may determine the unique $X$ (and hence the unique $R$) that achieves this $\hat{M}$ by the expression
\[X = 2\imath (\tilde{N}^{\flat})^{-1} (J\hat{M}-J\bar{M})\, \tilde{N}^{-1}.\blacksquare \]

As in the passive case, there is a continuum of choices for the cavity parameters, leading to different realizations of the system. We demonstrate this method with an illustrative example.
\begin{example}\label{example 4}
For the system of Example \ref{example 3}, the eigenvalue decomposition of $\mathcal{N}=N^{\flat}N$ is computed to be $\mathcal{N}=U D U^{-1}$, where $D=\diag(-2.8284,2.8284,-2.8284,2.8284)$ and
\[ U=\left(\begin{array}{rrrr}
-0.9074  &  0.3474  &  0.2038  &  0.1756\\
-0.1329  &  0.2965  &  0.4090  & -0.8908\\
0    &     0 &  -0.8629 &   0.4064\\
0.3987 &  -0.8896 &  -0.2159 &  -0.1027\\
\end{array}\right).\]
To the positive eigenvalue $\lambda^+=2.8284$, there correspond the eigenvectors $u_2$ and $u_4$ given by the second and fourth columns of $U$. We have that $\langle u_4,u_4\rangle_J >0$, and after normalization $u_4$ becomes $z^{+}=(0.2180,-1.1061,0.5046,-0.1275)^{\top}$. To the negative eigenvalue $\lambda^-=-2.8284$, there correspond the eigenvectors $u_1$ and $u_3$ given by the first and third columns of $U$. We have that $\langle u_1,u_1\rangle_J >0$, and after normalization $u_1$ becomes $z^{-}=(-1.0987,-0.1609,0,0.4827)^{\top}$. According to the proof of Lemma \ref{Bogoliubov SVD} \cite{gripet15},
\begin{eqnarray*}
&& W = \big[ [z^+ z^-]\, \Sigma\,[z^+ z^-]^{\#}\big] \\
&=&\left(\begin{array}{rrrr}
0.2180 &  -1.0987  &  0.5046   &      0\\
-1.1061 &   -0.1609  &   -0.1275 &   0.4827\\
0.5046   &    0  &  0.2180 &  -1.0987\\
-0.1275  &  0.4827 &  -1.1061 &  -0.1609\\
\end{array}\right).
\end{eqnarray*}
Since there are no zero eigenvalues,
\[\hat{N}=\bar{N}=
\left(\begin{array}{rrrr}
1.6818   &      0    &     0    &     0\\
0    &     0    &     0  &   1.6818\\
0    &     0  &  1.6818    &     0\\
0  &  1.6818   &      0    &     0\\
\end{array}\right), \]
and we can compute $V$ simply by
\begin{eqnarray*}
&& V= N\, W\, \hat{N}^{-1} \\
&=& \left(\begin{array}{rrrr}
-0.0576  & -1.0196  &  0.1834  & -0.0957\\
-1.0691  &  0.0164  &  0.3357  &  0.1749\\
0.1834  & -0.0957  &  -0.0576  & -1.0196\\
0.3357  &  0.1749 &  -1.0691  &   0.0164\\
\end{array}\right).
\end{eqnarray*}
The Hamiltonian of the reduced system should be equal to
\begin{eqnarray*}
&&\hat{M}=W^{\dag}M W \\
&=&
\left(\begin{array}{rrrr}
3.6444  &  1.0135  &  0.4429  & -3.3952\\
1.0135  &  4.3462  & -3.3952  & -1.7249\\
0.4429  & -3.3952  &  3.6444  &  1.0135\\
-3.3952 &  -1.7249  & 1.0135  &  4.3462\\
\end{array}\right).
\end{eqnarray*}
The reduced system can be implemented by the use of two cavities, one with a passive port (corresponding to $\lambda^+$), and one with an active port (corresponding to $\lambda^-$). Choosing the detuning of both cavities to be zero, makes the total Hamiltonian of their concatenation $M_{conc}=0_{4 \times 4}$. Also, we choose $\tilde{N}=I_4$.
Then, we compute
\begin{eqnarray*}
&& X = 2\imath (\tilde{N}^{\flat})^{-1} J\, (\hat{M}-M)\, \tilde{N}^{-1} \\
&=& \imath \left(\begin{array}{rrrr}
7.2889  &  2.0271  &  0.8858  &  -6.7904\\
2.0271  &  8.6924  & -6.7904  &  -3.4497\\
-0.8858 &   6.7904 &  -7.2889 &  -2.0271\\
6.7904  &  3.4497  &  -2.0271 &  -8.6924\\
\end{array}\right),
\end{eqnarray*}
from which the feedback gain $R$ is computed to be
\begin{eqnarray*}
&& R = (X-I)(X+I)^{-1} \\
&=& \,\,\, \left(\begin{array}{rrrr}
-0.3731  &  0.9082  & 0        & 0.0450 \\
0.9082   &  0.3125  & -0.0450  & 0      \\
0        &  0.0450  & -0.3731  & 0.9082 \\
-0.0450  &  0       & 0.9082   & 0.3125 \\
\end{array} \right) \\
&+& \imath \left(\begin{array}{rrrr}
7.8624 &  - 5.2659 & 7.4743 & - 5.8003\\
- 5.2659  &   4.4401 & - 5.8003 &  3.7042\\
- 7.4743  &   5.8003 & - 7.8624 &  5.2659\\
5.8003 &  - 3.7042 &  5.2659 & - 4.4401\\
\end{array} \right).\,\square
\end{eqnarray*}
Figure \ref{Active_Network_2} provides a graphical representation of the proposed implementation of the transfer function for this example.
\begin{figure}[!h]
\begin{center}
\scalebox{.3}{\includegraphics{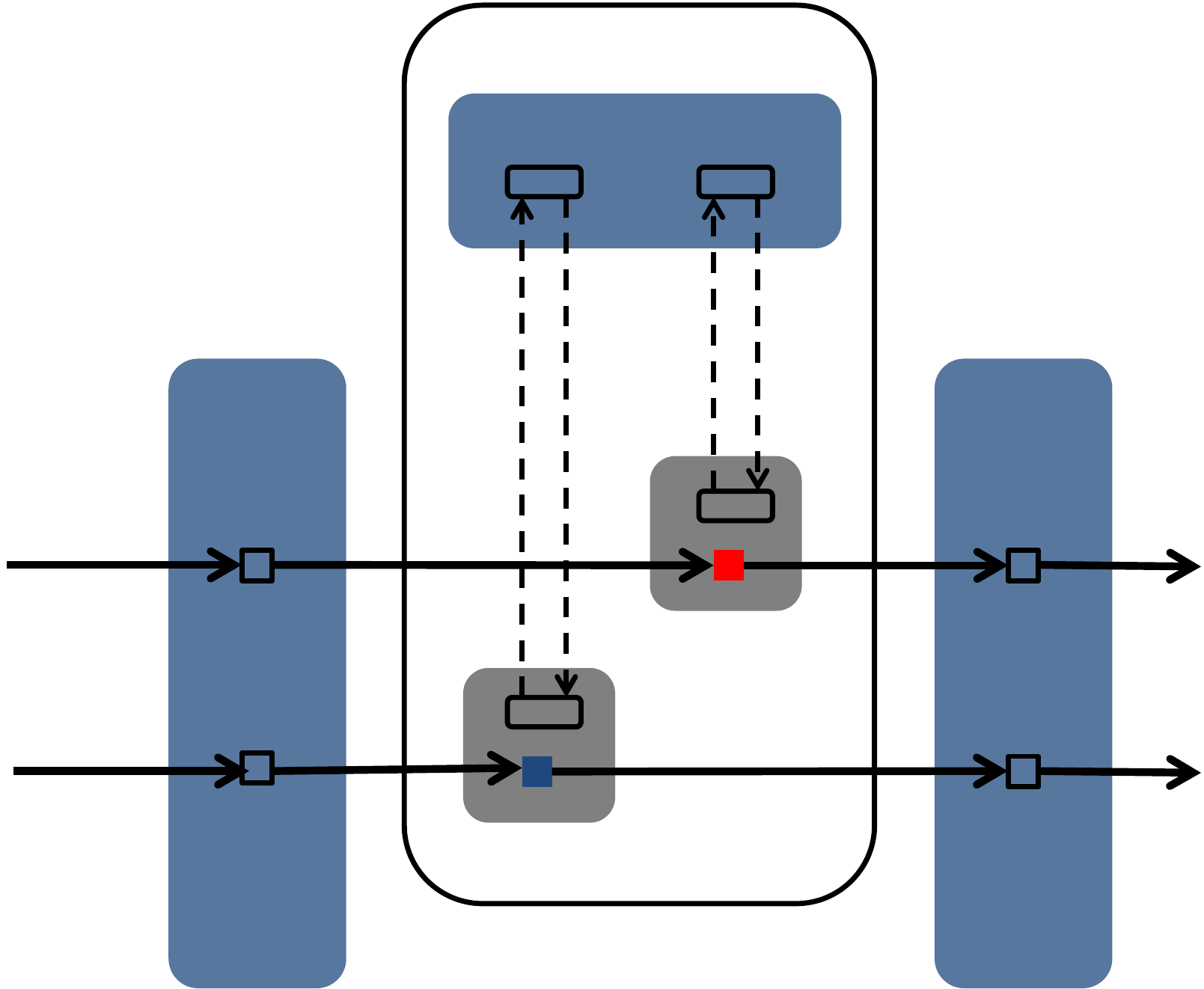}} \caption{Graphical representation of the proposed implementation of the transfer function of Example \ref{example 4}.} \label{Active_Network_2}
\end{center}
\end{figure}
\end{example}

\bibliographystyle{ieeetr}
\bibliography{C:/Users/Symeon/Documents/AAA/Work/Latex/MyBibliographies/Linear_Quantum_Stochastic_Systems,C:/Users/Symeon/Documents/AAA/Work/Latex/MyBibliographies/Books,C:/Users/Symeon/Documents/AAA/Work/Latex/MyBibliographies/My_papers}
\end{document}